\documentclass{emulateapj}
\usepackage{apjfonts}
\usepackage{multirow}
\usepackage{array}
\usepackage[usenames]{color}
\begin{document}
\submitted{ApJ accepted}
\newcommand{\comment}[1]{}
\newcommand{\risa}[1]{\textcolor{red}{(\bf #1)}}
\newcommand{\michael}[1]{\textcolor{blue}{(\bf #1)}}
\definecolor{purple}{RGB}{160,32,240}
\newcommand{\peter}[1]{\textcolor{purple}{(\bf #1)}}
\newcommand{\macc}{M_\mathrm{acc}}
\newcommand{\mpeak}{M_\mathrm{peak}}
\newcommand{\mnow}{M_\mathrm{now}}
\newcommand{\vacc}{v_\mathrm{acc}} 
\newcommand{\vpeak}{v_\mathrm{peak}} 
\newcommand{\vnow}{v^\mathrm{now}_\mathrm{max}}

\newcommand{\hinv}{h^{-1}}
\newcommand{\mpc}{\rm{Mpc}}
\newcommand{\hmpc}{$\hinv\mpc$}

\shortauthors{BEHROOZI ET AL}
\shorttitle{Gravitationally Consistent Halo Catalogs and Merger Trees}

\title{Gravitationally Consistent Halo Catalogs and Merger Trees for Precision Cosmology}

\author{Peter S. Behroozi, Risa H. Wechsler, Hao-Yi Wu}
\affil{Kavli Institute for Particle Astrophysics and Cosmology, 
Department of Physics, Stanford University; \\
Department of Particle Physics and Astrophysics,  SLAC National  Accelerator Laboratory; 
Stanford, CA 94305\\
{\tt behroozi@stanford.edu; rwechsler@stanford.edu}} 
 \author{Michael T. Busha}
  \affil{Institute for Theoretical Physics, University of Zurich, 8006 Zurich, Switzerland}
\author{Anatoly A. Klypin}
\affil{Astronomy Department, New Mexico State University, Las Cruces, NM, 88003}
\author{Joel R. Primack}
\affil{Department of Physics, University of California at Santa Cruz, Santa Cruz, CA 95064}

\begin{abstract}
  We present a new algorithm for generating merger trees and halo catalogs which explicitly ensures consistency of halo properties (mass, position, and velocity) across timesteps.  Our algorithm has demonstrated the ability to improve both the completeness (through detecting and inserting otherwise missing halos) and purity (through detecting and removing spurious objects) of both merger trees and halo catalogs.  In addition, our method is able to robustly measure the self-consistency of halo finders; it is the first to directly measure the uncertainties in halo positions, halo velocities, and the halo mass function for a given halo finder based on consistency between snapshots in cosmological simulations.  We use this algorithm to generate merger trees for two large simulations (Bolshoi and Consuelo) and evaluate two halo finders (\textsc{rockstar} and \textsc{bdm}).  We find that both the \textsc{rockstar} and \textsc{bdm} halo finders track halos extremely well; in both, the number of halos which do not have physically consistent progenitors is at the 1-2\% level across all halo masses.  Our code is publicly available at {\tt http://code.google.com/p/consistent-trees}.  Our trees and catalogs are publicly available at {\tt http://hipacc.ucsc.edu/Bolshoi/} .

\end{abstract}

\newcommand{\Mnfw}{M_\mathrm{NFW}}
\newcommand{\Msun}{M_{\odot}}
\newcommand{\mvir}{M_\mathrm{vir}}
\newcommand{\rvir}{R_\mathrm{vir}}
\newcommand{\vmax}{v_\mathrm{max}}
\newcommand{\vmac}{v_\mathrm{max}^\mathrm{acc}}
\newcommand{\mvac}{M_\mathrm{vir}^\mathrm{acc}}
\newcommand{\sfr}{\mathrm{SFR}}
\newcommand{\plotgrace}[1]{\includegraphics[width=\columnwidth,type=pdf,ext=.pdf,read=.pdf]{#1}}
\newcommand{\plotgraceflip}[1]{\includegraphics[width=\columnwidth,type=pdf,ext=.pdf,read=.pdf]{#1}}
\newcommand{\plotlargegrace}[1]{\includegraphics[width=2\columnwidth,type=pdf,ext=.pdf,read=.pdf]{#1}}
\newcommand{\plotlargegraceflip}[1]{\includegraphics[width=2\columnwidth,type=pdf,ext=.pdf,read=.pdf]{#1}}
\newcommand{\plotminigrace}[1]{\includegraphics[width=0.5\columnwidth,type=pdf,ext=.pdf,read=.pdf]{#1}}
\newcommand{\plotmicrograce}[1]{\includegraphics[width=0.25\columnwidth,type=pdf,ext=.pdf,read=.pdf]{#1}}
\newcommand{\plotsmallgrace}[1]{\includegraphics[width=0.66\columnwidth,type=pdf,ext=.pdf,read=.pdf]{#1}}

\section{Introduction}

Over the past few decades, dark matter simulations have demonstrated increasing usefulness for validating theories of cosmology, for understanding systematic biases in observations, and for constraining galaxy and large-scale structure formation.  In coming years, the rapid expansion of observational data coming from ground and space based surveys, including CANDELS, GAMA, BOSS, DES,  
Herschel, Pan-STARRS, BigBOSS, eROSITA, Planck, JWST, and LSST will mean that simulations will become even more important for modeling and understanding the detailed evolution of the cosmos.  This wealth of data means that cosmological and galaxy properties soon will be measured to a new standard of precision; however, none of this will increase the accuracy of current cosmological constraints without a concordant increase in the quality of simulations and our ability to model the systematic biases inherent in the observations.

Two of the principal outputs of dark matter simulations are halo catalogs and merger trees; namely, information about the deep potential wells where galaxies are expected to reside, and a history of the mergers and growth of these potential wells.  Derived properties of these outputs, such as the halo mass function and auto-correlation function, must be understood at the one-percent  and five-percent level, respectively, in order to use the full constraining power of future surveys for, e.g., dark energy \citep{Wu10}.  Similar levels of accuracy are required to be able to distinguish between different values of the primordial non-gaussianity parameter $f_\mathrm{NL}$ \citep{Pillepich10}.  In addition to raw accuracy, models of galaxy formation (e.g., semi-empirical abundance matching and semi-analytical models) depend on the physical consistency of catalogs and merger trees, in the sense that they require physically reasonable halo growth histories and dynamically-plausible mergers to accurately model the build-up of galaxy properties (e.g., stellar mass, luminosity, metallicity, dust) over time \citep{Benson11}.

To date, while simulations largely agree on the final dark matter distribution, few comprehensive reviews have been performed to determine which combinations of halo finders and merger tree codes produce the most accurate results \citep[see, however, ][]{Knebe11}.  In part, this is because cosmological halos have complicated structure; depending on the particle distribution, it may be difficult to tell which halo finder is ``better'' or ``worse'' except by using a tedious and subjective examination by hand.  On the other hand, with comparisons performed on more clinical test cases, such as halos generated with perfect NFW profiles, it is difficult to know how the comparison results will translate to the messier world of cosmological halos.  Furthermore, percent-level understanding of the halo mass function requires not only that the halo finder should function well in the common cases, but that it should also be robust against even the most extremely misshapen halos.

This paper avoids the problem of {\em defining} accuracy on a cosmological simulation, and instead seeks to provide some clarity on this issue with a different approach.  As noted earlier, a necessary (although not sufficient) precondition for accuracy is physical consistency.  In most simulations, halo properties are expected to evolve slowly relative to the rate at which simulation timesteps are saved.  As such, by comparing halos across several timesteps, it becomes possible to analyze not only which halo finders most consistently determine halo properties, but also, which halo finders result in the most reasonable evolution of halos, based on, e.g., their mass accretion, positions, and velocities.

Our choice for the halo properties to compare across timesteps (i.e., halo mass, circular velocity, position, and velocity) implies that we must calculate the gravitational evolution of halos (as distinct from particles) across timesteps.  This approach requires somewhat more effort than simpler approaches, but it nonetheless has several unique advantages.  The most obvious one is that we can do extended tests on the physical consistency of halos, and thus, we may make \textit{quantifiable} estimates of, e.g., the current accuracy of halo mass functions.  Just as usefully, the approach allows us to repair halo catalogs and merger trees when inconsistencies are found --- e.g., when a halo disappears for a few timesteps in the halo catalogs, we can regenerate its expected properties by gravitational evolution from the surrounding timesteps.  Here, we use this approach to generate two sets of halo catalogs and merger trees for two large simulations (Bolshoi and Consuelo), which use different simulation codes (\textsc{art} and \textsc{gadget-2}, respectively) from halos found using the \textsc{rockstar} halo finder \citep{Rockstar}.  In addition, we generate merger trees for Bolshoi using the \textsc{bdm} halo finder \citep{Klypin99,Bolshoi}, which allows us to compare the consistency of different halo finders on the same simulation.

This paper is divided into sections as follows.  First, we discuss limitations of merger trees and halo finding in the literature which can cause inconsistencies in \S \ref{s:history}.  Next, we present details including cosmology assumptions for the two main simulations (Bolshoi and Consuelo) and the two halo finders (\textsc{rockstar} and \textsc{bdm}) for which we calculate merger trees in \S \ref{s:sims}.  Then, we describe the methods we use to predict halo locations and velocities across timesteps (i.e., the gravitational evolution methods) in \S \ref{s:grav_desc}.  We discuss the metrics and methods for repairing merger trees and halo catalogs in \S \ref{s:grav_repair} and present quantitative tests for these methods in \S \ref{s:tests}.  Finally, we summarize our conclusions in \S \ref{s:conclusions}.

\section{Current Methods and Current Issues with Merger Trees}

\label{s:history}

\subsection{A Brief Overview of Current Methods}

In order to determine the likely locations of galaxies, dark matter simulations are postprocessed to find gravitationally self-bound groups of particles (\textit{host halos}, also called central halos), which may themselves contain subgroups of self-bound particles (\textit{subhalos}, sometimes also called satellite halos).  Traditionally, merger trees have been generated by tracking particles in identified halos from one timestep to another.  In most approaches, a halo at one timestep (a \textit{progenitor}) is linked to a halo at the next timestep (the \textit{descendant}) if the majority of the particles in the progenitor end up in the descendant \cite[e.g.,][]{ASOHF10,Zhao09,Fakhouri08,Li07merger,Nagashima05,Helly03,Hatton03,Wechsler02,Tormen98,Roukema97,Cole94}.

Generally, enhancements to this basic method have assigned more weight to the trajectories of the most-bound particles in each halo; this implies better continuity for the galaxy at the center of the dark matter potential well.  Such methods range from ensuring that the most-bound particle is located within the descendant halo \cite[e.g.,][]{vdBosch02,Kauffmann99}, to creating merger trees based on the trajectories of a fraction of the most-bound particles \cite[e.g.,][]{Cole08,Harker06,Okamoto00}, and more complicated continuous weighting metrics \cite[e.g.,][]{Fakhouri10,DeLucia07,Springel05Mill}.  For specialized purposes, such as calculating smooth accretion vs.\ substructure accretion, some studies have examined splitting halos such that all the particles in each resulting group end up in the same descendant \citep{Genel10,Genel09}.

In some more recent implementations \citep{Springel05Mill,AllgoodThesis,Allgood06,Wetzel09, Wetzel10,Bolshoi} attempts have been made to extend progenitor identification beyond particle-based merger trees to ensure halo continuity; however many of these attempts have focused on robust substructure tracking using a subset of halo properties.  Here we extend this approach 
to consider a wide range of properties for both halos and subhalos, including halo mass, maximum circular velocity ($\vmax$), position, and bulk velocity.   As halo finders have known imperfections---e.g., problems finding halos near the resolution limit and problems resolving substructure near the centers of host halos---these imperfections translate into problems with accretion histories in purely particle-based merger trees.  Just as importantly, halo finders can have \textit{unknown} imperfections which particle-based merger trees cannot reveal.\footnote{Indeed, several previously unknown halo finding issues with both \textsc{bdm} and \textsc{rockstar} were identified and fixed as a result of developing this algorithm.}  Hence, it is the desire to fix both the known and the unknown problems with merger trees which has motivated us to attempt more advanced methods for tree construction.
We present a sampling of a few known problems (which affect all halo finders) in the following section (\S \ref{s:consistency}).

\subsection{Consistency Problems}

\label{s:consistency}

Particle-based halo finders and merger tree algorithms have been used very successfully in matching galaxy populations from very high redshifts ($z\sim 8$) to the present day for massive clusters to dwarf satellites \citep{Behroozi2012}; some recent examples of the relevant simulations include \cite{Bolshoi,MICE,MillII,VLII,Aquarius,Springel05Mill}.  At the same time, there are several instances in which problems can appear that do not satisfy future requirements for high-precision halo catalogs and trees:

\begin{enumerate}
\item A subhalo may not be identified during one or more timesteps where its orbit passes close to the center of a larger halo.  As a result, in the merger tree, it would be classified as merged with the larger halo (which would receive most of its particles), but then when it reappears, the subhalo would be identified as a new halo with no progenitors.
\item In the less extreme case where the halo finder identifies a subhalo close to the center of a larger halo but where many of the particles are mis-identified as being associated with the host halo, there would be a similar result: the merger tree would record a false merger and the sudden appearance of a new halo with no progenitors.  Approaches which track only the most-bound particles would result in fewer such cases in the merger tree, but the halo properties (e.g., halo mass, $v_\mathrm{max}$, etc.) would remain incorrect for those timesteps.
\item For halos whose particles are distributed among many other halos at the next timestep, the fate of the halo's galaxy is not clear---depending on which particles end up in other halos, the galaxy could either be disrupted into the intracluster light, or it could end up in one of the halos which received the most-bound particles.  This has partial overlap with case (1), as two merging halos may be mis-identified as a single halo during a close approach and then be identified as multiple halos at the next timestep.
\item The opposite effect may also happen---a subhalo passing close to the center of a larger halo may erroneously be assigned particles from the host halo, leading either to a spuriously large subhalo or a duplicate of the host halo.
\item Halos just on the threshold of identification (e.g., because of low particle numbers) may appear and disappear several times over successive timesteps, leading to false mergers, halos with no descendants, or simply a bias against low-mass halos, depending on the merger tree implementation.
\end{enumerate}

Each of these problems leads to systematic biases in recovering subhalo properties.  Under the assumption that galaxies reside in subhalos as well as host halos \citep[e.g.,][]{conroy:06,cw-08,Lu12,Behroozi2012}, many of these problems would affect precision comparisons with observations.  For example, issues with subhalos would reduce the number of close galaxy pairs predicted by the simulation.  Issues with halos undergoing major mergers would result in systematic miscounting of the halo mass function.  This is particularly important at the massive end of the mass function, where mergers are common to the present data and accurate constraints are required for precision cosmology.  Finally, issues with halos not having correct progenitor tracks would result in either incorrect modeling (in terms of semi-analytic galaxy models) or incorrect matching (in terms of the abundance matching approach) of galaxies in halos, making it more difficult to compare galaxy catalogs between simulations and observations.  Many of these problems have been addressed to varying degrees in the literature previously, in particular those issues relating to robust tracking of substructure, see for example \cite{Wechsler02, Springel05, Faltenbacher05, Allgood06, Harker06, Wetzel09, Tweed09}.  Here we attempt to address all of these issues systematically and simultaneously.

\section{The Simulations and Halo Finders}

In this paper, we present merger trees for two large $\Lambda$CDM dark matter simulations (Bolshoi and Consuelo).  These simulations demonstrate the applicability of our method to two different simulation codes (\textsc{art} and \textsc{gadget}-2, respectively), as well as two different halo finders (\textsc{rockstar} and \textsc{bdm}), each with different strengths and weaknesses.  We describe Bolshoi in \S \ref{s:bolshoi}, Consuelo in \S \ref{s:consuelo}, the \textsc{rockstar} algorithm in \S \ref{s:rockstar}, the \textsc{bdm} algorithm in \S \ref{s:bdm}, and we describe the initial particle-based merger trees for Bolshoi and Consuelo in \S \ref{s:particletrees}.  We have also conducted some limited comparisons with the \textsc{subfind} algorithm \citep{Springel01} on a subregion of Bolshoi in Appendix \ref{a:subfind}.  Throughout this paper, we assume that halo masses are calculated as spherical overdensities including contributions from any substructure.

\label{s:sims}

\subsection{Bolshoi}

\label{s:bolshoi}

We present merger trees for a new high-resolution simulation, \textit{Bolshoi},
described in detail in \cite{Bolshoi}.  Bolshoi follows a comoving, periodic box
with side length 250 $h^{-1}$ Mpc with 2048$^3$ ($\approx 8.6 \times 10^9$) particles
from redshift 80 to the present day.  Its exquisite mass resolution 
($1.9 \times 10^8$ $\Msun$ per particle) and force resolution (1 $h^{-1}$ kpc) make it
ideal for studying intrinsic properties, clustering, and evolution of halos from $10^{10}$
 $\Msun$ (e.g., satellites of the Milky Way) to the largest clusters in the universe
 ($10^{15}$ $\Msun$).  Bolshoi was run as a collisionless dark matter simulation with
the Adaptive Refinement Tree Code \citep[\textsc{art}; ][]{kravtsov_etal:97,kravtsov_klypin:99}
assuming a flat, $\Lambda$CDM cosmology ($\Omega_M = 0.27$, $\Omega_\Lambda = 0.73$,
$h = 0.7$, $\sigma_8 = 0.82$, and $n_s = 0.95$).  These cosmological parameters
are consistent with results from both WMAP5 \citep{wmap5} and the latest WMAP7+BAO+H$_0$ results \citep{wmap7}.  Our merger trees are constructed using 180 output snapshots of the simulation, and contain
a total of nearly 3 billion halos.

\subsection{Consuelo}

\label{s:consuelo}

We also present merger trees for a second simulation, \textit{Consuelo}, which is one box out of a suite of 200 taken from the Large Suite of Dark Matter Simulations (McBride et al, in preparation).\footnote{LasDamas Project, {\tt http://lss.phy.vanderbilt.edu/lasdamas/}} Consuelo covers a larger volume (420 $h^{-1}$ Mpc on a side) with fewer particles ($1400^3$) making it ideal for studies of cosmic variance in high-redshift surveys.  The mass resolution per particle ($2.7 \times 10^9$ $\Msun$) implies a completeness limit close to $2\times10^{11}$ $\Msun$ for halo masses; the reduced force resolution (softening length of 8 $h^{-1}$ kpc) as compared to Bolshoi implies a reduced ability to track subhalos within the virial radius of a larger halo.  Consuelo was run as a collisionless dark matter simulation using \textsc{gadget}-2 \citep{Springel05}, with a flat, $\Lambda$CDM cosmology ($\Omega_M = 0.25$, $\Omega_\Lambda = 0.75$, $h = 0.7$, $\sigma_8 = 0.8$, and $n_s = 1.0$) which is similar to the WMAP5 best-fit cosmology \citep{wmap5}.
Our default merger trees for this simulation are constructed using 100 output snapshots
and contain a total of approximately 500 million halos in all timesteps.  We present results from Consuelo largely in Appendix \ref{a:consuelo} to streamline the content of the main body of the paper.

\subsection{The \textsc{rockstar} Halo Finder}

\label{s:rockstar}

For the main results in this paper, both the Bolshoi and Consuelo simulations were analyzed using the \textsc{rockstar} halo finder.  The \textsc{rockstar} algorithm is a newly-developed phase-space temporal (7D) halo finder designed for increased consistency and accuracy of halo properties, especially for subhalos and major mergers \citep{Rockstar}.  The method first divides the simulation volume into 3D friends-of-friends groups with a large linking length ($b=0.28$) for easy parallel analysis.  For each group, particle positions and velocities are normalized by the group position and velocity dispersions, giving a natural phase-space metric.  Then, the algorithm adaptively chooses a phase-space linking length such that 70\% of the group's particles are linked together into subgroups.  This process repeats for each subgroup---renormalization, a new linking-length, and a new level of substructure calculated---until a full hierarchy of particle subgroups is created.  Seed halos are then placed in the densest subgroups, and particles are assigned hierarchically to the closest seed halo in phase space (see \citealt{Rockstar} for full details).  Finally, once particles have been assigned to halos, unbound particles are removed and halo properties (positions, velocities, spherical masses, radii, spins, etc.) are calculated.  If halos at the previous snapshot are available, they are used to determine the host halo / subhalo relationships in cases (such as major mergers) where they are ambiguous.

\subsection{The \textsc{bdm} Halo Finder}

\label{s:bdm}

The basic technique of the \textsc{bdm} halo finder is described in \cite{Klypin97}; a more detailed description is given in \citet{Riebe11}, and tests and comparisons with other codes are presented in \citet{Knebe11}. The code uses a spherical 3D overdensity algorithm to identify halos and subhalos.  It starts by finding the density for each individual particle; the density is defined using a top-hat filter with a given number of particles $N_{\rm filter}$, which typically is $N_{\rm filter}=20$. The code finds all density maxima, and for each maximum it finds a sphere containing a given overdensity mass $M_\Delta=(4\pi/3)\Delta\rho_{\rm cr}R^3_\Delta$, where $\rho_{\rm cr}$ is the critical density of the Universe and $\Delta$ is the specified overdensity.

Among all overlapping spheres the code finds the one that has the
deepest gravitational potential. The density maximum corresponding to
this sphere is treated as the center of a distinct halo. Thus, by
construction, a center of a distinct halo cannot be inside the radius
of another one. However, peripheral regions can still partially
overlap, if the distance between centers is less than the sum of halo
radii.  The radius and mass of a distinct
halo depend on whether the halo overlaps or not with other distinct
halos. The code takes the largest halo and identifies all other
distinct halos inside a spherical shell with distances $R=(1-2)R_{\rm
 center}$ from the large host halo, where $R_{\rm center}$ is the
radius of the largest halo. For each halo selected within this shell,
the code finds two radii. The first is the distance $R_{\rm big}$ to
the surface of the large halo: $R_{\rm big} =R-R_{\rm center}$. The
second is the distance $R_{\rm max}$ to the nearest density maximum in
the shell with the inner radius $\min( R_{\rm big},R_\Delta)$ and the
outer radius $\max( R_{\rm big},R_\Delta)$ from the center of the
selected halo. If there are no density maxima within that range, then
$R_{\rm max}= R_\Delta$. The radius of the selected halo is the
maximum of $R_{\rm big}$ and $R_{\rm max}$. Once all halos around the
large halo are processed, the next largest halo is taken from the list
of distinct halos and the procedure is applied again. This setup is
designed to make a smooth transition of properties of small halos when
they fall into a larger halo and become subhalos.

The bulk velocity of either a distinct halo or a subhalo is defined as the
average velocity of the 100 most bound particles of that halo or by
all particles, if the number of particles is less than 100. The number
100 is a compromise between the desire to use only the central
(sub)halo region for the bulk velocity and the noise level.

The gravitational potential is found by first finding the mass in
spherical shells and then by integration of the mass profile. The
binning is done in $\log$ radius with a very small bin size of
$\Delta\log(R) =0.01$.

Centers of
subhalos can only be found among density maxima, but not all density
maxima are subhalos.  An important construct for finding
subhalos are barrier points: a subhalo radius cannot be larger than
the distance to the nearest barrier point times a numerical tuning factor
called an overshoot factor $f_{\rm over}\approx 1.1-1.5$.
The subhalo radius can be smaller than this distance. Barrier
points are centers of previously identified (sub)halos. For the first
subhalo, the barrier point is the center of the distinct halo. For the
second subhalo, it is the first barrier point and the center of the
first subhalo, and so on. The radius of a subhalo is the minimum of
(a) the distance to the nearest barrier point times $f_{\rm over}$ and
(b) the distance to its most remote bound particle.

\subsection{Particle-Based Merger Trees}

\label{s:particletrees}

As part of the algorithm process for generating gravitationally consistent trees, we computed simple particle-based merger trees for both Bolshoi and Consuelo.  The algorithm assigns a descendant to a halo based on which halo at the next timestep receives the largest fraction of the halo's particles (excluding substructure).  In principle, this method is sufficient to correctly predict the vast majority of halo descendants (although of course it cannot identify cases where a halo should never have existed, or where a halo was missed).  An algorithm based on a fixed fraction of the most-bound particles would be superior in cases where a subhalo is undergoing rapid stripping (e.g., when it loses more than half of its particles to its host); however, the algorithm we present in this paper can very effectively correct for such cases.

Indeed, for the \textsc{bdm} analysis of Bolshoi, only the 250 most-bound particles were available; as such, we created initial merger trees based on only these 250 particles for each halo.  For small halos (<2000 particles), this approach worked well; however, there were more issues for large halos and halos undergoing major mergers.  Specifically, the use of incomplete particle information resulted in a small fraction ($\approx 0.2-0.5\%$ between timesteps) of halos without descendants and a large fraction (10\%) of spurious links (see \S \ref{s:grav_links}).  By comparison, the fraction of spurious links in the \textsc{rockstar} particle trees was between 1-3\%, depending on redshift.  The initial \textsc{bdm} tree thus represents a more challenging set of initial conditions, but it serves as an ideal proving ground for the efficacy of our algorithm.

\section{Gravitational Halo Evolution Equations}
\label{s:grav_desc}

\subsection{Overview}

To solve the problems identified in \S \ref{s:consistency}, it is necessary to enforce consistency in the halo catalog across timesteps.  No matter how well-written the halo finder is, there will always be halos which (for example) cross the threshold of detection in one timestep and then disappear in the next.  It is impossible to tell whether those cases are statistical fluctuations or not based only on the information available at a single timestep---otherwise, presumably, appropriate logic could be added to the halo finder to account for them.  Thus, the presence or absence of a halo in adjacent timesteps lends otherwise unavailable evidence which helps determine whether the halo should be present in the current timestep.

Knowing the positions, velocities, and mass profiles of halos at one timestep, we may use the laws of gravity and inertia to predict their properties at adjacent timesteps.  By comparing the predicted halo catalogs with the actual ones, and by calculating the deviations from the predictions, we can immediately tell whether the halo finder has missed or misidentified halos.  The approach taken herein is straightforward and effective when the halo catalogs have already been calculated.\footnote{A future paper may explore integration with halo finders in order to improve halo identification \textit{in situ} as the halo catalogs are being generated.}  We detail the equations used in this model in the next two sections; first for predicting bulk halo motion (\S \ref{s:grav_ev}), and second for predicting tidal disruption of halos into a more massive host (\S \ref{s:tidal_mergers}).  Detailed tests of the model accuracy are presented in later sections of the paper (\S \ref{s:grav_tests} and \S \ref{s:tidal_tests}).

\subsection{Gravitational Evolution for Predicting Most-Massive Progenitors}

\label{s:grav_ev}

In predicting halo motion between successive timesteps, we make a few simplifying assumptions.  First, we assume that the kinematics of dark matter halos are principally affected only by the positions and mass profiles of other dark matter halos in the simulation.  While this assumption breaks down at the very highest redshifts, it remains remarkably accurate for halos at currently-observable redshifts (out to at least $z\sim 10$; see \S \ref{s:grav_tests}).  To additionally reduce the complexity of our code, we approximate individual halo mass distributions by fitting spherical NFW profiles \citep{NFW97}.  Thus, each halo is fully described by a position vector, a velocity vector, a scale radius ($r_s$), and a characteristic density ($\rho_0$).  Again, while this assumption is incorrect in detail, it nonetheless remains remarkably accurate for tracking halo motion between consecutive timesteps (\S \ref{s:grav_tests}).

Using the positions, velocities, and halo profile information for halos at one timestep, we may predict the positions and velocities of their most-massive progenitors.  We apply Newtonian gravity to halos embedded in the standard FLRW expanding coordinate system.  In the usual formula, the force between two halos would be:\footnote{Note that as dark matter halos are extended and non-rigid, the notion of the ``force'' between two halos is somewhat ambiguous.  In this case, we define the force on a halo to be the acceleration of its center times the virial mass of the halo so that the expected equation $F = ma$ still holds.}
\begin{equation}
|F_{1\to2}| = \frac{G M_1 M_2}{r^2}.
\end{equation}

Where $r$ is the distance between halo centers.  We approximate the mass $M_1$ as the dark matter bound to the first halo within a radius $r$ of its center, subtracting the mass of subhalos (if any).  Once the NFW parameters are determined for the halo in question, this may be calculated by integrating the NFW profile out to the desired radius:
\begin{eqnarray}
\Mnfw(r, r_s, \rho_0) & = & 4\pi\rho_0 \int_0^r \frac{r_s}{r}\left(1+\frac{r}{r_s}\right)^{-2} r^2dr \nonumber \\
&=& 4\pi \rho_0 r_s^3\left[\ln\left(1+\frac{r}{r_s}\right) - \frac{r}{r+r_s}\right]
\end{eqnarray}
In addition, we note that subhalos have a steadily decreasing gravitational influence on the bulk motion of the host as they approach the host's center, due to their overlapping mass distributions.  Because subhalos are often much smaller than their host, we find it sufficient to introduce a softening length of a fraction ($\epsilon$) of the virial radius to avoid this problem.\footnote{We take the virial mass ($M_\mathrm{vir}$) and radius ($r_\mathrm{vir}$) of a halo to be defined in terms of the virial spherical overdensity ($\Delta_\mathrm{vir}$) with respect to the mean background density as given by \cite{mvir_conv}.  I.e., $\Delta_\mathrm{vir} =
  (18\pi^2 + 82 x - 39 x^2)/(1+x)$;
  $x=(1+\rho_\Lambda(z)/\rho_M(z))^{-1}-1$}
  In major mergers, this results in a somewhat inaccurate force law; however, in these cases, the dominant form of momentum transfer is via particles changing halo membership, which is substantially more difficult to model correctly.  Nonetheless, even though we do not model this process, the average velocity errors are still quite reasonable, as demonstrated in \S \ref{s:tests}.  In our tests, the choice of $\epsilon$ had little impact on how well positions and velocities were predicted, with the default $\epsilon = 0.2$ performing marginally (<5\% better errors) better on average than $\epsilon = 1$.
 
  Thus, the full force equation becomes
\begin{equation}
|F_{1\to2}| = \frac{G \Mnfw(r, r_{s,1}, \rho_{0,1}) M_\mathrm{vir,2}}{r^2 + (\epsilon r_\mathrm{vir, 2})^2},
\end{equation}
and the contribution to the acceleration of the second halo is
\begin{equation}
\Delta\vec{a}_{2} = -\frac{G \Mnfw(r, r_{s,1}, \rho_{0,1})}{r^2 + (\epsilon r_\mathrm{vir, 2})^2}\hat{r}.
\label{e:accel}
\end{equation}
We do not find it necessary to introduce additional terms for dynamical friction, even on the order of timesteps which are 300Myr, as it remains a subdominant source of error in terms of calculating halo velocities.

While the most straightforward way to predict halo locations would be to evaluate forces between every pair of halos, this would require $\mathcal{O}(n^2)$ time to compute.  This becomes prohibitive for today's best simulations (where the number of halos is $n \gg 10^6$), so we describe an approach to restrict the computation time to roughly $\mathcal{O}(n \ln(n))$.

  For a desired accuracy in halo velocities ($\Delta v$) and a given timestep length ($\Delta t$), we need only calculate accelerations ($\Delta a$) for nearby halos; namely, those which satisfy $\Delta a \gtrsim \Delta v/\Delta t$.  To first order, given Eqn.~\ref{e:accel}, a halo will accelerate nearby halos only as a function of the separation distance:
\begin{equation}
\Delta a \sim \frac{G M_\mathrm{vir}}{r^2}
\end{equation}
We may then define a cutoff distance beyond which we need not calculate the gravitational effects of a given halo:
\begin{equation}
r_\mathrm{cutoff}(M_\mathrm{vir}) = \sqrt{\frac{GM_\mathrm{vir}}{\Delta v / \Delta t}},
\end{equation}
as the given halo will not affect the velocities of other halos beyond $r_\mathrm{cutoff}$ by more than the desired velocity accuracy.  Using binary space partitioning (BSP) trees to efficiently access halos by location,\footnote{BSP trees are a superset of $kd$-trees.  In the latter approach, the size of the refinement volumes is generally fixed at a given refinement level.  In our approach, the size of the refinement volumes shrinks to exactly cover the enclosed
elements, yielding slightly better memory usage and access times.} we thus limit the amount of work we have to do for each halo to be proportional to the number of halos within a distance $r_\mathrm{cutoff}$.  Profiles of our implementation suggest that the work required to build and access the BSP trees is much larger than the work required to calculate gravitational forces, which results in a runtime which scales as $\mathcal{O}(n \ln(n))$.

We find that setting a velocity tolerance of $\Delta v = 5$ km s$^{-1}$ and using leapfrog integration is more than adequate for the purpose of tracking halos; this results in a systematic drift for predicted halo positions of at most 5 pc Myr$^{-1}$ (see also \S \ref{s:grav_tests}).

\subsection{Tidal Forces and Mergers}

\label{s:tidal_mergers}

In the case where a halo does not match up to the predicted locations of potential descendants, it may either be a halo which briefly fluctuates above the threshold for halo detection, or it may be a halo which merges into a larger halo at the next time step.  Some indication of which of these two fates occurred may be obtained via an estimate of the tidal field at the center of the halo in question, i.e., the spatial derivative of the acceleration field (see Eq.\ \ref{e:accel}).  To leading order, the tidal field exerted by one halo on another will be:\footnote{The coefficient of proportionality varies from 6-10, under the assumption of NFW profiles, and depends only weakly on the distance between the two halos.}
\begin{equation}
\label{e:tidal_field}
\frac{d\vec{a}}{dr} \propto |T| \equiv \frac{G M_{NFW}(r, r_{s,1}, \rho_{0,1})}{r^3}
\end{equation}

In particular, we find that a simple threshold cut on the value of $|T|$ reliably separates halos which could reasonably undergo mergers from those that could not (see \S \ref{s:tidal_tests} for validation).

\section{A Gravitationally Consistent Method for Repairing Halo Catalogs and Merger Trees}

\label{s:grav_repair}

\subsection{Overview}

In this section, we assume the existence of halo catalogs for every output timestep of a dark matter simulation.  Particle-based merger trees are required for calibrating the metric for calculating likely progenitors; however, as such, they do not need to be computed for the entire volume of the box, nor do they need to be especially robust.  As we demonstrate at the end of \S \ref{s:grav_links}, using merger trees based on only the 250 most-bound particles gives identical final results in our approach as using merger trees based on the trajectories of all particles in each halo.

The consistency method that we adopt is by no means the only one.  Nonetheless, our gravitational method has several unique advantages:
\begin{enumerate}
\item Missing halos may be fully reconstructed, with consistent positions, velocities, and halo properties, along with quantifiable error estimates for the reconstructions.
\item Merger tree links are assigned a natural likelihood estimate.  Particle-based links between halos which are too far apart in position, velocity, or mass may then be cut and reconnected to more likely candidates.
\item The resolution limit of the simulation---i.e., the particle number below which halo properties are no longer reliably calculated---becomes explicitly quantifiable in terms of the errors induced in the position and velocity of halos.
\item The method subjects the general reliability of the halo finder to an independent check; different halo finders may then be compared on an even footing to evaluate how self-consistently they recover halo properties across multiple timesteps.
\item The method provides a clean way to distinguish between subhalos which are tidally disrupted at the next timestep and subhalos which are instead lost by the halo finder, which would otherwise have an identical particle-based merger tree.
\end{enumerate}

We first discuss some basic methodology in terms of linking progenitors and descendants (\S \ref{s:halo_links}) and then discuss the two stages of our algorithm (\S \ref{s:grav_links} and \S \ref{s:cleanup_links}).

\subsection{Linking Progenitors and Descendants}

\begin{figure*}[!th]
\Large
\begin{center}
\setlength{\extrarowheight}{2ex}
\begin{tabular}{m{300pt}@{\hspace{4ex}}m{250pt}}
\textsc{Halo Merger Tree Algorithm}\\
\textbf{1.} Identify halo descendants using a traditional particle algorithm. 
&  \includegraphics[scale=0.25,type=pdf,ext=.pdf,read=.pdf]{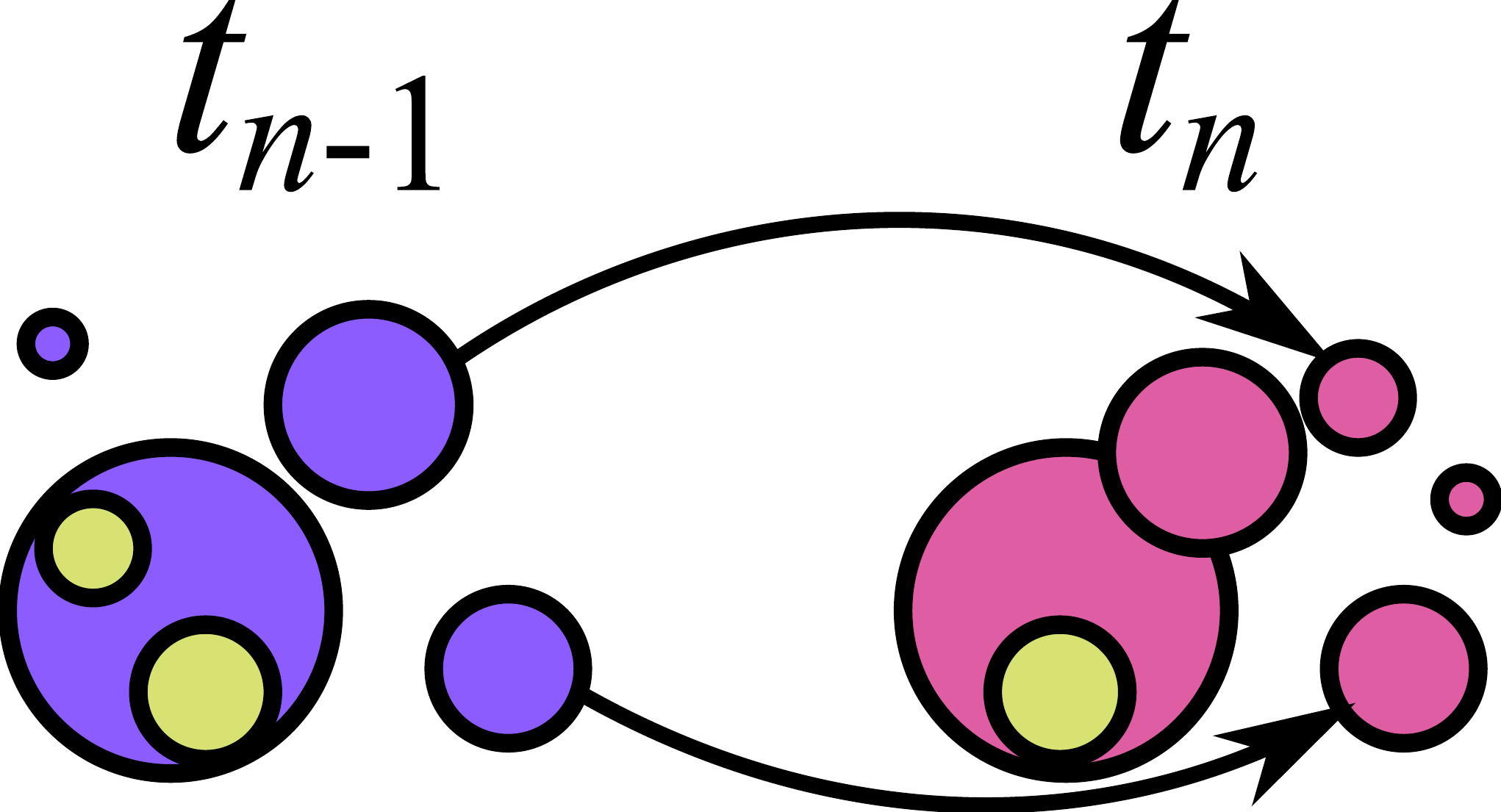} \\
\textbf{2.} Gravitationally evolve the positions and velocities of all halos at the current timestep back in time to identify their most likely positions at the previous timestep.
&  \includegraphics[scale=0.25,type=pdf,ext=.pdf,read=.pdf]{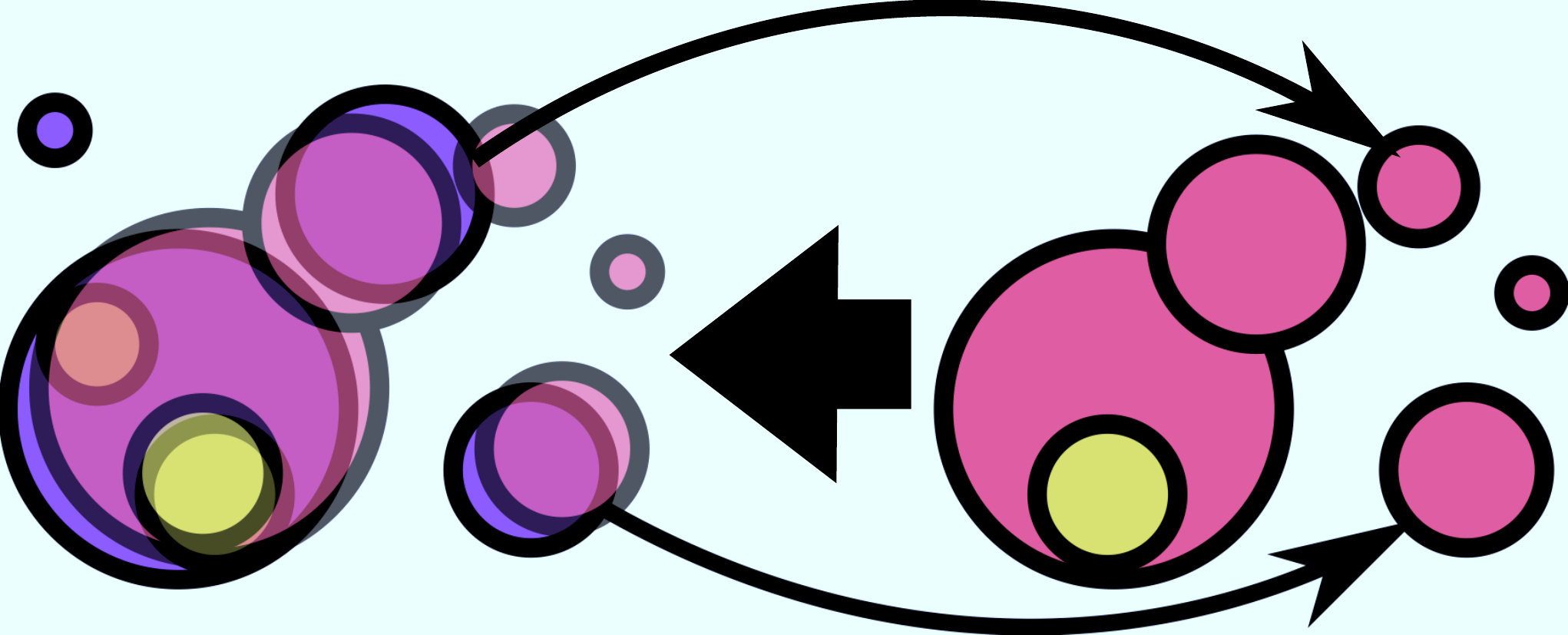} \\
\textbf{3.} Based on predicted progenitor halos in step (2), cut ties to spurious descendants.
&  \includegraphics[scale=0.25,type=pdf,ext=.pdf,read=.pdf]{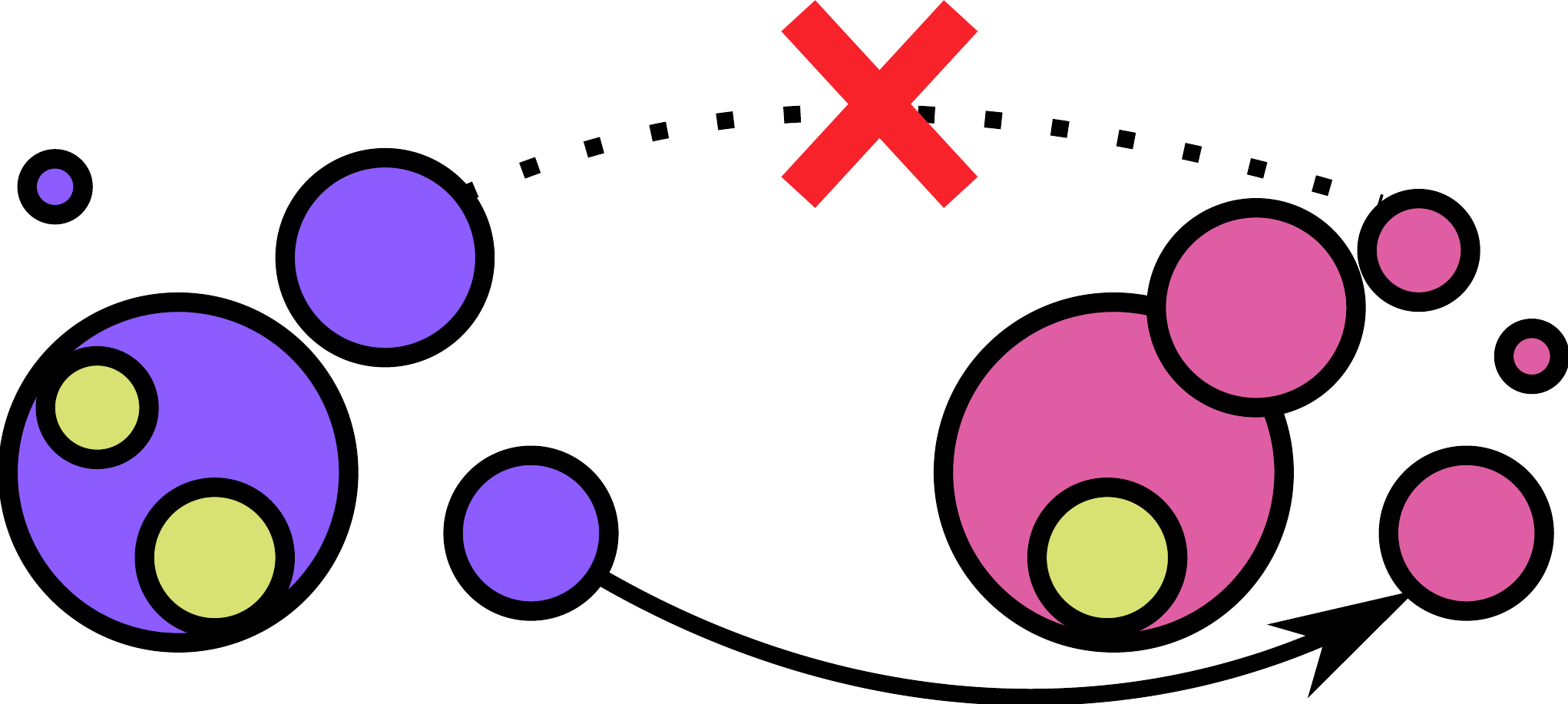} \\
\textbf{4.} Create links for halos with likely progenitors at the previous timestep for cases in which step (2) has identified a good match.
&  \includegraphics[scale=0.25,type=pdf,ext=.pdf,read=.pdf]{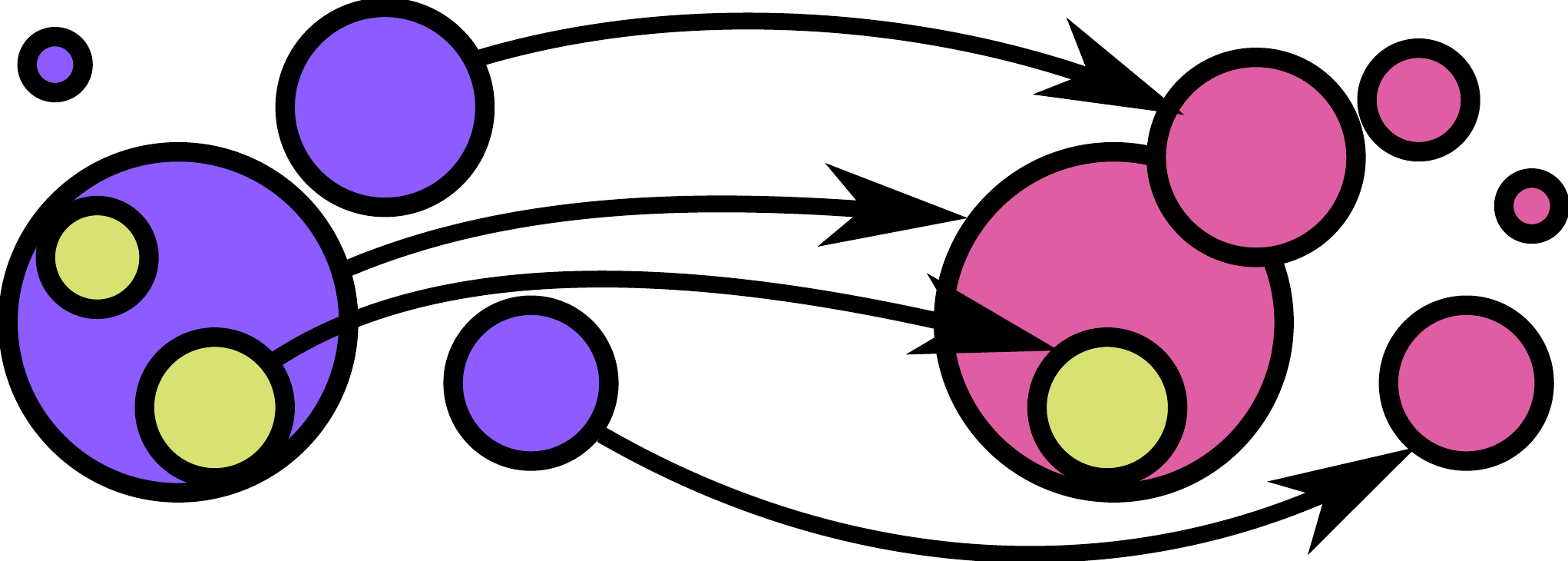} \\
\textbf{5.} For halos in the current timestep without likely progenitors, create a new halo at the previous timestep with position and velocity given by the evolution in step (2).  Remove any such halos generated from previous rounds if they have had no real progenitors for several timesteps.
&  \includegraphics[scale=0.28,type=pdf,ext=.pdf,read=.pdf]{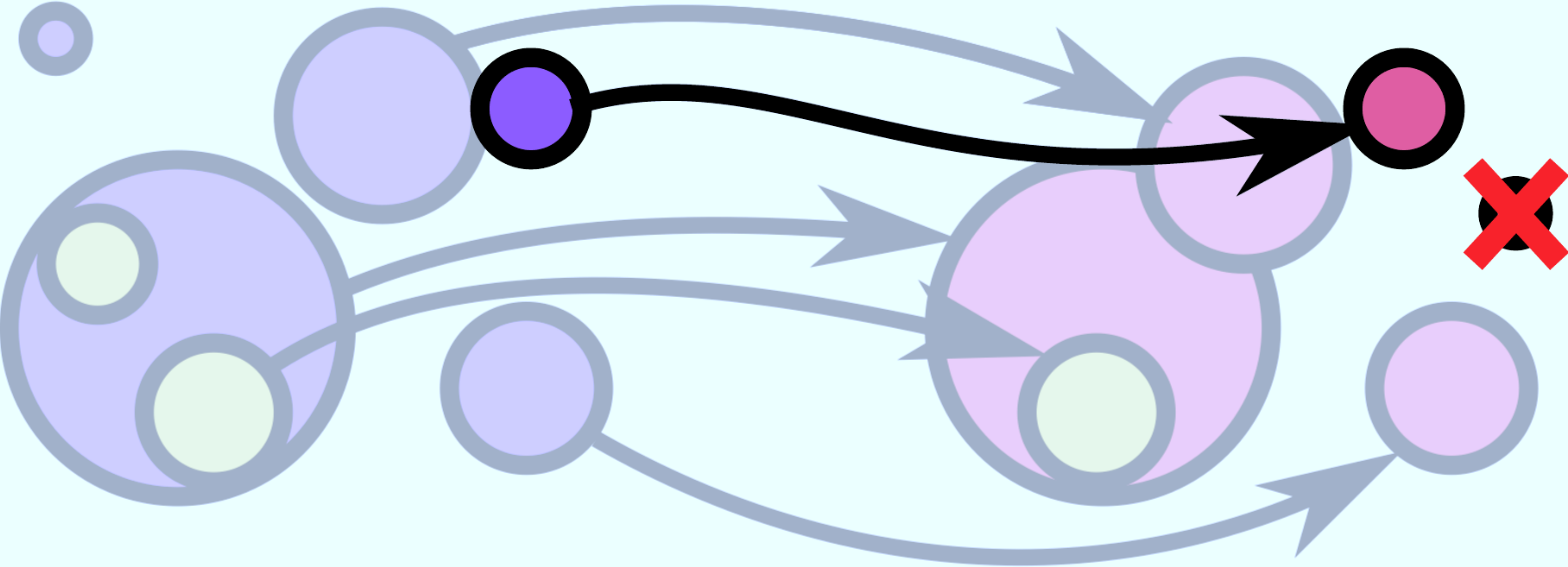} \\
\textbf{6.} For halos in the previous timesteps which have no descendants, assume that a merger occurred into the halo exerting the strongest tidal field across it at the previous timestep.  If a halo with no descendant is too far removed from other halos to experience a significant tidal field, assume that it is a statistical fluctuation and remove it from the tree and catalogs.
&  \includegraphics[scale=0.28,type=pdf,ext=.pdf,read=.pdf]{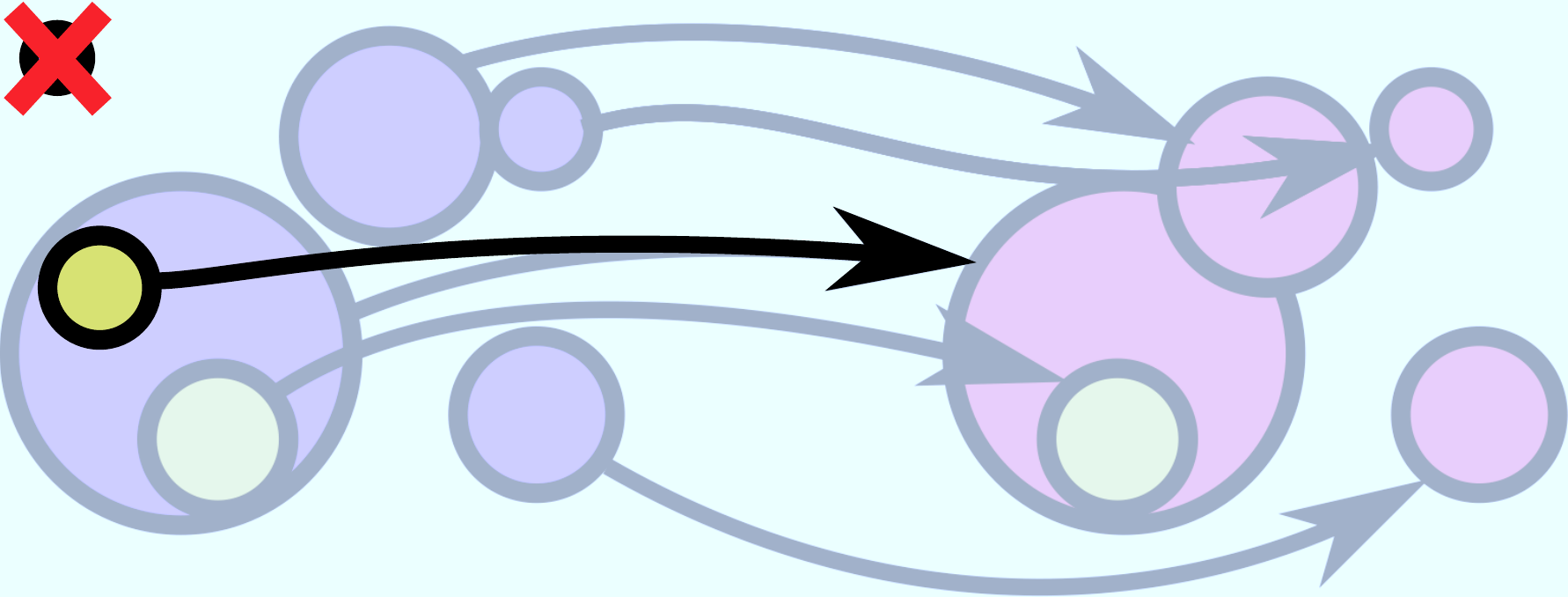} \\
\end{tabular}
\setlength{\extrarowheight}{0ex}
\normalsize
\caption{A visual summary of the first stage of the merger tree algorithm.}
\label{f:summary}
\end{center}
\end{figure*}

\label{s:halo_links}

Underlying our approach for repairing merger trees is the observation that in $\Lambda$CDM, halos do not spontaneously appear with large masses; instead, they are built up from smooth accretion and mergers of smaller halos.  This implies that every halo has \textit{at least} one progenitor at the previous timestep (although the mass of the progenitor may be too small for the halo finder to recover it).  If we trace a halo backwards to its expected location at the previous timestep but do not find a progenitor there, we may conclude that the halo catalog is incomplete.  If we find a match in the halo catalogs after tracing the halo backwards for a few more timesteps, we may interpolate the intrinsic halo properties (e.g., halo mass, $\vmax$, etc.) between the timesteps and assign positions and velocities based on the best estimates of the gravitational evolution algorithm (\S \ref{s:grav_ev}).  However, if we do not find a good match, we may either conclude that the halo is small enough to have just formed or that the halo is a spurious detection and should be removed.

The same is not true if we were to perform the gravitational evolution in the opposite direction (i.e., forward).  Since it is common for halos to merge together, the absence of a unique descendant is not immediate evidence for an inconsistency in the halo catalogs.  Knowledge of the tidal forces does help with this ambiguity, but many subhalos temporarily disappear just when they pass by the center of a larger halo---just where the tidal forces are the strongest.  For that reason, more robust statements about the consistency of the halo catalogs can be made if the gravitational evolution is performed \textit{backwards}, i.e., from each timestep back to the next earlier one.

Across timesteps, many halo properties are expected to change slowly (e.g., $\vmax$, $\mvir$, $\rvir$, and angular momentum) or predictably (e.g., position and velocity).  These properties may then be used to tell whether a given halo has a reasonable progenitor in the catalog.  To calibrate what is considered ``reasonable,'' we use particle-based merger trees to determine the accuracy of our predictions for progenitor properties as compared to the actual progenitor properties in the trees.  The characteristic errors in predicting position ($\tau_x$), velocity ($\tau_v$), and $\vmax$ ($\tau_{vmax}$) yield a natural distance metric, $d$.\footnote{We expect $\mvir$ and $\rvir$ to be highly correlated with $\vmax$, except in the case of subhalos, where they are harder to predict and consequently yield less accessible information than $\vmax$.  Thus, we exclude them from direct consideration in our metric.  In a future revision of our code, we may add support for angular momentum comparisons between halos, but we do not do so at present because this property is not yet reported by all halo finders.}  In particular, if we denote the expected progenitor properties with a subscript $e$ and those of a candidate progenitor by subscript $c$, we have:
\begin{equation}
\label{e:metric}
d(e, c)  =  \sqrt{\frac{|\vec{x}_e - \vec{x}_c|^2}{2\tau_x^2} + \frac{|\vec{v}_e - \vec{v}_c|^2}{2\tau_v^2} + \frac{\log_{10}\left(\frac{v_{\mathrm{max}, e}}{ v_{\mathrm{max}, c}}\right)^2}{2\tau_{vmax}^2}}
\end{equation}

This gives a natural ranking for candidate progenitors, as well as a natural way to supply a threshold (i.e., maximum acceptable value for the metric) for physical consistency.

Of course, there are many ways of determining characteristic scales for errors.  In our code, we choose to take the sum of the average error and the standard deviation of the error for position and velocity (e.g., $\tau_x = \langle |\Delta x| \rangle + \sigma_{|\Delta x|}$) because this yields increased robustness against unusual probability distributions.  Yet as might be expected, in practice, we find that both the average error and the standard deviation of the error are always within a factor of two of each other.  We find additionally that $\tau_x$ and $\tau_v$ have a dependence on halo mass; we account for this by binning halos by 0.25 dex in mass and then calculating $\tau_x$ and $\tau_v$ separately for each bin.\footnote{This results in an ambiguity in Eq.\ \ref{e:metric}---e.g., should one use $\tau_x(M_{\mathrm{vir}, e})$ or $\tau_x(M_{\mathrm{vir}, c})$?  Arguments can be made for and against both choices, with neither one being obviously superior.  Computationally, however, it is easier to choose (as we do) to use the expected properties (e.g., $\tau_x(M_{\mathrm{vir}, e})$).}  For estimating $v_{\mathrm{max}, e}$ in Eq.\ \ref{e:metric}, we can calculate the fractional change in $\vmax$ over each timestep in the same mass bins.  We then may take $\tau_{vmax}$ to be the standard deviation of the logarithmic change in $\vmax$ across timesteps as a function of mass (i.e., $\sigma_{\log_{10}(\Delta \vmax)}(\mvir)$).

Note that defining the metric in this way has the unique advantage that it is not necessary to calculate particle-based merger trees for all halos.  If the trees are available for even a small portion of the simulation, that is sufficient to calculate the distance metric (that is, the values of $\tau_x$, $\tau_v$, and $\tau_{vmax}$) which applies to the entire volume.  This feature makes our algorithm suitable even when particle IDs are not stored for all particles or when particle IDs are not consistent across the entire volume (as in tiled simulations).

\subsection{Stage One: Fixing Links and Filling in Missing Halos}

\label{s:grav_links}

As described in the previous section, every correctly-identified halo must have a progenitor at the previous timestep, except for those halos whose progenitors are below the mass-resolution limit of the simulation.  As such, we begin by evolving the halos at one timestep ($t_n$) backwards to the previous timestep ($t_{n-1}$) to predict the properties of the expected most-massive progenitor.  As explained in the previous section, these predictions in combination with the particle-based merger trees allow us to calculate a metric $d(e, c)$ to evaluate the likelihood that a candidate halo $c$ at timestep $t_n$ is the most-massive progenitor of a halo $e$ at timestep $t_{n-1}$---i.e., that the connection or \emph{link} between $c$ and $e$ is physically reasonable.

Once calculation of the metric is complete, we break all potentially problematic links in the particle-based merger trees.  These include:
\begin{enumerate}
\item All links where the metric $d(e, c)$ is above some predetermined threshold $d_{break}$.  We choose $d_{break} = 3.2$, which results in 1-2\% of all particle-based links to be broken in the Bolshoi simulation.
\item All links where the progenitor is not the most-massive progenitor of the descendant halo.  This affects 2--3\% of all halo links in the Bolshoi simulation.  These links can be problematic in two cases---either a) the descendant halo was not identified by the halo finder, or b) the descendant halo identified in the particle-based merger trees is a host halo containing the actual descendant halo.  Hence, it is important to consider all options for descendants of these halos before concluding that they represent tidal mergers into the most massive host.
\item All links where the most-massive progenitor is beyond a predetermined ratio in $\mvir$ or $\vmax$ from the descendant halo.  We choose $M_\mathrm{vir,break}$ = 0.5 dex and $v_\mathrm{max,break}$ = 0.15 dex; this affects 2\% of all particle-based links in the Bolshoi simulation at $z=0$ for \textsc{rockstar}, increasing to 4\% (depending on halo mass) at very high redshifts where the mass accretion rate is higher.  In contrast, for \textsc{bdm}, the limited number of particles used to determine links (250) results in 10-20\% of links at high halo mass ($M>10^{13} \Msun$) being clearly spurious in this way, with a direct correlation to the timestep length.  Such links are almost always cases where the most-massive progenitor was misidentified or mis-linked in the particle-based merger tree---and hence, those halos may have different descendants as mentioned in the previous item.
\end{enumerate}

Then, for all halos at $t_n$ without progenitors, we scan for potential progenitors among all the halos without descendants at $t_{n-1}$.  To do so, we rank potential progenitors by likelihood according to the metric $d(e,c)$; if any potential progenitors are found within a predetermined threshold $d_{match}$, the one with the highest likelihood is assigned as the most-massive progenitor of the halo in question.  We set $d_{match} = 15$; for all of our tested combinations of halo finders and simulation codes, this still restricts progenitors to be well within the virial radius of their descendant.

With \textsc{bdm}, we find that this procedure is not sufficient for some outlying cases.  In particular, we find that the halo-finding algorithm used in \textsc{bdm} has occasional trouble locating centers of massive objects.  In cases where multiple dense peaks are present within an overdense region (as is the case with major mergers), \textsc{bdm} may switch between those peaks in successive timesteps when it attempts to determine halo properties.  As such, massive halos may in rare cases ($<3$\% of merger tree links) switch to a new location up to a virial radius away for dozens of timesteps or more \citep[see also, e.g., discussion in][]{Wetzel09}.  These cases are obviously not physical---but even so, it is impossible to call one of the centers more ``correct'' than the other without a careful phase-space analysis.  Because \textsc{bdm} and many other popular halo finders do not find halos in phase space, we have chosen to explicitly allow an exception for such cases in our merger trees.  As such, for those halos which still do not have physically acceptable progenitors, we allow a progenitor to be matched at the previous timestep if a) it is within the virial radius of the descendant halo, and b) if its $\vmax$ is within $v_\mathrm{max,break} = 0.15$ dex of the descendant halo.

Even so, some halos at $t_n$ will still have no progenitors; for these halos, two options remain.  Either the progenitor is missing from the halo catalog at the previous timestep, or the progenitor has fallen below the mass-completeness limit of the simulation.  Determining which option is correct requires analysis of earlier timesteps---however, for large simulations like Bolshoi, only a few timesteps may be able to fit into memory simultaneously.  To allow for a more flexible analysis, we create a placeholder halo, called a \emph{phantom halo}, in the halo catalog at timestep $t_{n-1}$ for each halo remaining at $t_n$ without a progenitor.  Phantom halos may be created at several successive timesteps to allow for cases in which the halo finder loses track of a halo for multiple timesteps; this latter case most often occurs for major mergers.  However, to avoid spurious links between accidentally coincident true and phantom halos, we cease tracking phantom halos beyond a predetermined number of timesteps $t_{phant}$. For Bolshoi, we find that tracking phantom halos for up to four timesteps is sufficient to patch over the vast majority of cases for missing halos.

For halos at $t_{n-1}$ which still have no descendants, there are also two options.  Either they are not the most-massive progenitor of their descendant (i.e., they are tidal mergers), or they are spurious fluctuations in the halo catalogs.  We use the formula in Eq.\ \ref{e:tidal_field} to discriminate between these two cases.  In particular, we find that a tidal acceleration field below $|T|$ = 0.3-0.4 km s$^{-1}$ Myr$^{-1}$ comoving Mpc$^{-1}$ is a robust indicator that a tidal merger is extremely unlikely (see \S \ref{s:tidal_tests}).  As such, halos above that threshold are assigned descendants according to the halo exerting the largest tidal field; halos below that threshold are deleted from the catalogs.  This method agrees in over 95\% of cases with the original particle-based merger trees (see \S \ref{s:tidal_tests}), the remaining cases being those where subhalos were incorrectly merged into their host in the particle-based trees.

A graphic summary of the most important steps of this algorithm is shown in Fig.\ \ref{f:summary}.  As compared to the raw particle-based merger trees for the \textsc{bdm} halo finder on Bolshoi (which used only 250 particles to track mergers) this stage of the algorithm, 10-20\% of links at each timestep need repairs for halos with $M > 10^{13} \Msun$ and 5\% are repaired for halos with $M < 10^{13} \Msun$; these are largely halos where the progenitor was clearly mis-identified in the particle-based trees.  By comparison, in the particle-based trees for \textsc{rockstar} (which used all halo particles to track mergers), only 2--3\% of links at each timestep are changed.

\begin{table*}
\caption{Summary of algorithm parameters}
\label{t:parameters}
\begin{center}
\begin{tabular}{rcm{300pt}l}
\hline\\[-1.7ex]
\hline
Variable & Chosen Value & Description & Section\\
\hline
$\epsilon$ & 0.2 & Softening length (in units of the host virial radius) for subhalos' gravitational influence on their host. & \S \ref{s:grav_ev}\\
$d_{break}$ & 3.2 & Threshold for breaking links in the particle-based merger trees according to the distance metric. & \S \ref{s:grav_links}\\
$d_{match}$ & 15 & Threshold for considering a proposed link acceptable according to the distance metric. & \S \ref{s:grav_links}\\ 
$M_{vir,break}$ & 0.5 dex & Threshold for breaking most-massive progenitor links if the mass ratio of progenitor and descendant exceeds this value. & \S \ref{s:grav_links}\\
$v_{max,break}$ & 0.15 dex & Threshold for breaking most-massive progenitor links if the $\vmax$ ratio of progenitor and descendant exceeds this value. & \S \ref{s:grav_links}\\
$a_{tidal}$ & 0.4 km s$^{-1}$ Myr$^{-1}$ cmvg Mpc$^{-1}$ & Threshold for the tidal field to consider a tidal merger physically acceptable. & \S \ref{s:grav_links}\\
$f_{phant}$ & 25\% & Threshold of acceptability for the fraction of time a halo track may contain phantom halos. & \S \ref{s:cleanup_links}\\
$t_{phant}$ & 4 & Number of timesteps to calculate expected phantom halo locations. & \S \ref{s:grav_links}\\
$t_{tracked}$ & 5 & Minimum number of timesteps a halo can exist in the catalogs in order to be considered physical. & \S \ref{s:cleanup_links}\\
$t_{tracked,subs}$ & 10 & Minimum number of timesteps a subhalo can exist in the catalogs in order to be considered physical. & \S \ref{s:cleanup_links}\\
\hline
\end{tabular}
\end{center}
\end{table*}

\subsection{Stage Two: Cleaning Up Halo Tracks}

\label{s:cleanup_links}

The main effect of the previous stage is to create new halos where there are gaps in the halo catalogs.  However, comparatively few halos are removed---only those which have no obvious descendant at the next timestep.  Thus far, we have only been concerned with the physical validity of individual halos, as opposed to full halo \textit{tracks}---that is to say, the lineage of most-massive progenitors for a given halo.  Checking the validity of halo tracks extends the phase-and-mass-space checks of the previous section with temporal checks---e.g., it allows us to remove halos if they only appear for a few timesteps in the catalogs.  To complete the removal of spuriously-detected halos, we detect and remove three types of problematic halo tracks:

\begin{enumerate}
\item Halos whose lineage of most-massive progenitors contains more than a fraction $f_{phant}$ of phantom halos.  Halos with too many phantom progenitors are usually those which are on the bare threshold of detection.  For more massive halos which meet this criterion, it is extremely unlikely that they represent valid detections; the vast majority of those that do are subhalos passing close to the center of larger host halos which have been incorrectly assigned particles from the host.  We choose $f_{phant} = 25\%$, which removes about 0.1\% of halos at all timesteps in the Bolshoi simulation for \textsc{rockstar} and 0.3\% of halos at all timesteps for \textsc{bdm}.
\item Halos which are tracked for fewer than $t_{tracked}$ timesteps.  Again, for massive halos, it is extremely unlikely that they represent valid detections if they are not tracked beyond a few timesteps.  However, for smaller halos and earlier redshifts, setting this threshold too low can lead to removal of legitimate halo tracks.  We find that a value of $t_{tracked}=5$ represents a good compromise, removing 0.2--0.5\% of all halos across all timesteps in Bolshoi for both halo finders without adversely affecting halo accretion at early times.\footnote{A future version of the merger tree code will allow all such conditions to be specified in terms of a length of time, rather than a number of timesteps.}
\item Subhalos whose tracks do not extend outside of the virial radius of their host and which are tracked for fewer than $t_{tracked, subs}$ timesteps.  Ordinarily, one might expect all such halos to be spurious, but it is conceivable that a halo might form outside the virial radius of a host and be accreted in between timesteps if the timesteps are sufficiently far apart.  We find that a threshold of $t_{tracked, subs}=10$ removes an additional 0.1--0.4\% of halos at all timesteps in the Bolshoi simulation for \textsc{bdm} and 0.1\% for \textsc{rockstar}.
\end{enumerate}

We note that the default parameters are chosen to be fairly lax: Bolshoi has 180 timesteps, so it should be expected that more massive halos should be tracked for a large fraction of this number.  This is in fact the case for our merger trees (see \S \ref{s:tracking_tests}).  Nonetheless, as opinions may differ for what constitutes ``physical'' values for the number of timesteps tracked and the fraction of phantom halos, we only remove halo tracks which have the most obvious inconsistencies; the output format of our merger trees is such that users may easily implement more stringent tests depending on their needs.

During the clean-up stage, we also finalize halo properties for surviving phantom halos.  These remaining phantoms connect a real progenitor halo to a real descendant halo over one or more timesteps when the halo is missing or inconsistent in the halo catalogs.  The positions and velocities of the phantom halos are taken from the gravitational evolution algorithm in \S \ref{s:grav_ev}.  The masses and almost all other properties are taken as linear combinations of the properties of the progenitor and descendant halo.  The only exceptions are properties which scale as mass to the one-third power (e.g., the halo radius and $v_\mathrm{max}$); for these properties, their third power is linearly interpolated instead so as to remain consistent with the halo mass interpolation.  Some caution is necessary in interpreting properties of phantom halos: for example, a halo could truly lose mass at one timestep (thereby dropping below the detection threshold) and then gain mass at the next.  This would result in the interpolated properties overestimating the mass and size of the halo at the missing timestep; for that reason, we clearly mark all phantom halos in our catalogs so that it is easy to tell if they have an impact in any given calculation.

We summarize all the parameters used in our algorithm in Table \ref{t:parameters}.  While the number of algorithm parameters may seem large at first glance, this is a reflection of the large number of physical sanity tests which our approach enables in terms of excluding unphysical halos.  The fact that some halos fail each of the sanity tests for reasonable fiducial values of the parameters suggests that (at least with current halo finders) the integrity of the merger trees would otherwise be compromised by unphysical halo tracks.

\begin{figure}[!t]
\plotgrace{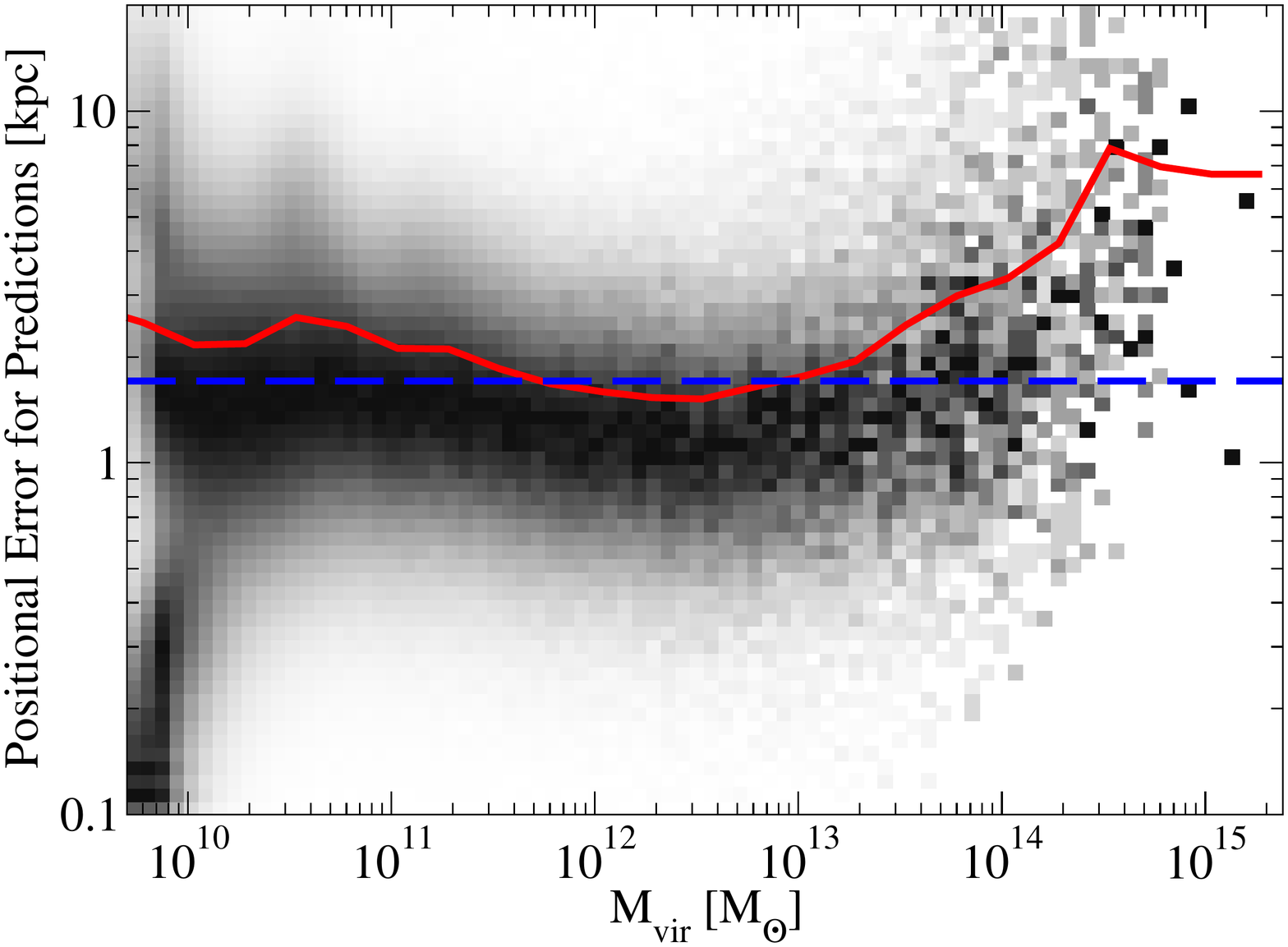}\\[-5ex]
\plotgrace{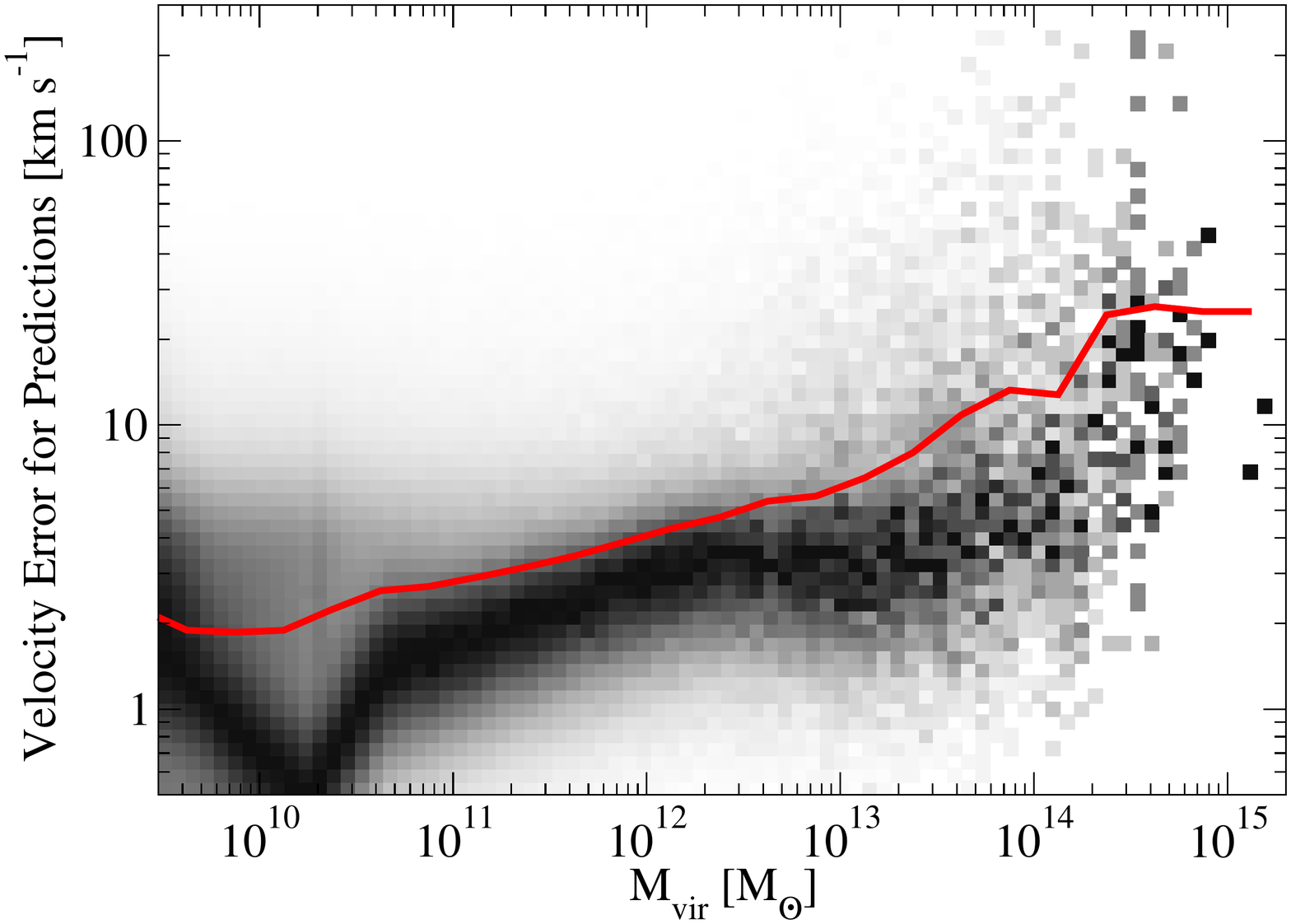}
\caption{Conditional density plots of the errors between the expected positions and velocities of halos (obtained via gravitational evolution from the subsequent timestep) and the actual values in the halo catalog as a function of halo mass.  These plots show the errors for one timestep lasting 42 Myr at $z=0$, using the Bolshoi simulation and the \textsc{rockstar} halo finder.  Superimposed over the density plots are red lines showing the linear average of the errors;  blue dashed line in the positional error plot (top) indicates the force resolution.
The bifurcation in positional errors at low masses is due to shot noise in locating halo centers; see text.  The linear average is significantly offset from the median for velocity errors due to a long tail in the error distribution.  The ``V''-shaped feature in the error distribution is due to how \textsc{rockstar} calculates velocities for halos with low particle numbers; see text.}
\label{errors}
\end{figure}

\begin{figure}[!h]
  \plotgrace{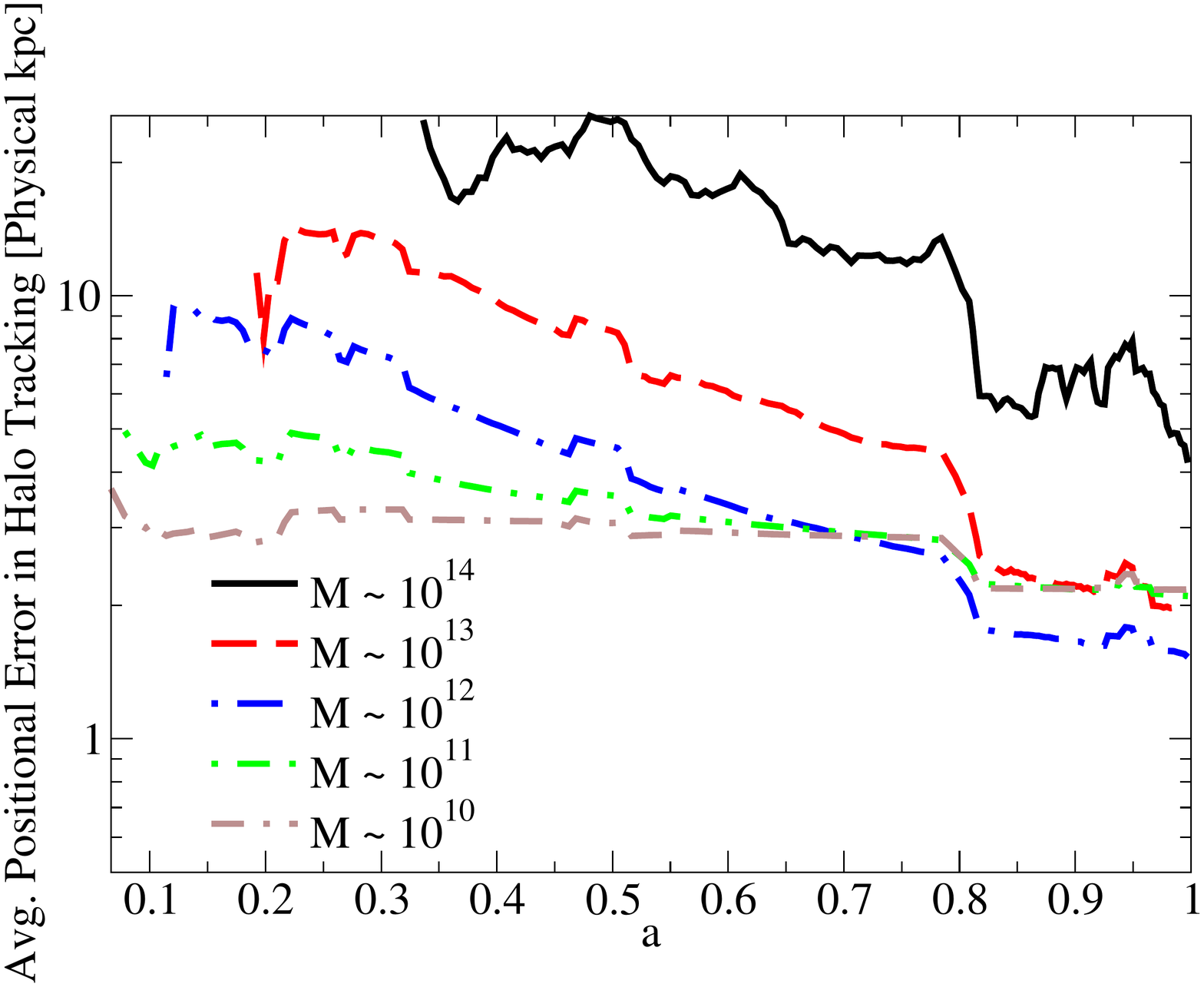}\\[-5ex]
  \plotgrace{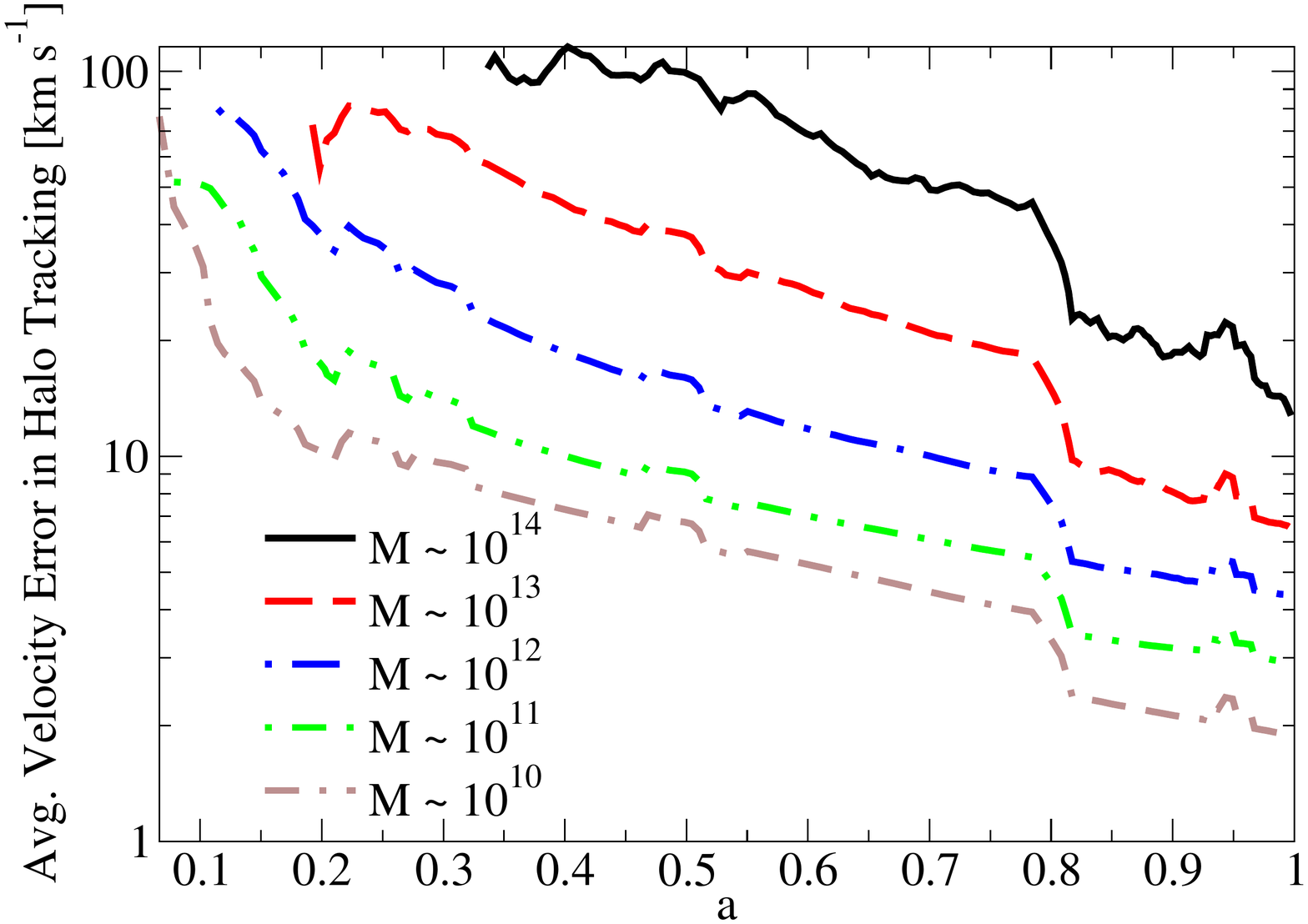} \caption{Comparison of errors at different timesteps for the \textsc{rockstar} halo finder on Bolshoi.  Positional errors (in real distance, as opposed to comoving distance) appear to be mostly independent of redshift, whereas velocity errors do not.  The velocity errors at later redshifts are reduced due to the reduced merger frequency.  For the most massive halos ($M\sim 10^{14}\Msun$), there is a clear break in the positional errors at $a \sim 0.8$; this is because the timestep length doubles for $a<0.8$, which suggests that velocity errors are largely responsible for the resulting positional errors at these masses.  All halo masses ($M$) are in units of $\Msun$.}
\label{error_comp}
\end{figure}

\begin{figure}[!h]
\vspace{-5ex}
\plotgrace{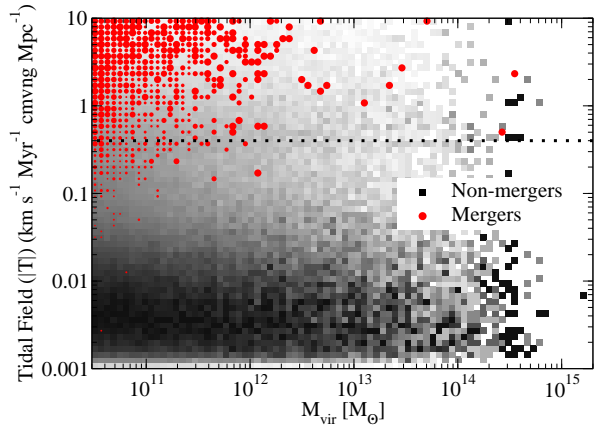}
\caption{A conditional density plot of the tidal force acting on \textsc{rockstar} halos in Bolshoi at $z=0.01$.  The grey density plot shows the tidal force acting on most-massive progenitors (i.e., halos which do not tidally merge), and the red density/size plot shows the tidal force acting on tidally merging halos (as identified in the particle-based merger trees).  The tidal force is expressed in terms of differential acceleration (km s$^{-1}$ Myr$^{-1}$) per unit distance (comoving Mpc); the dotted line represents our classification threshold for physically-acceptable tidal mergers.}
\label{tidal}
\end{figure}

\begin{figure}[!h]
\vspace{-5ex}
\plotgrace{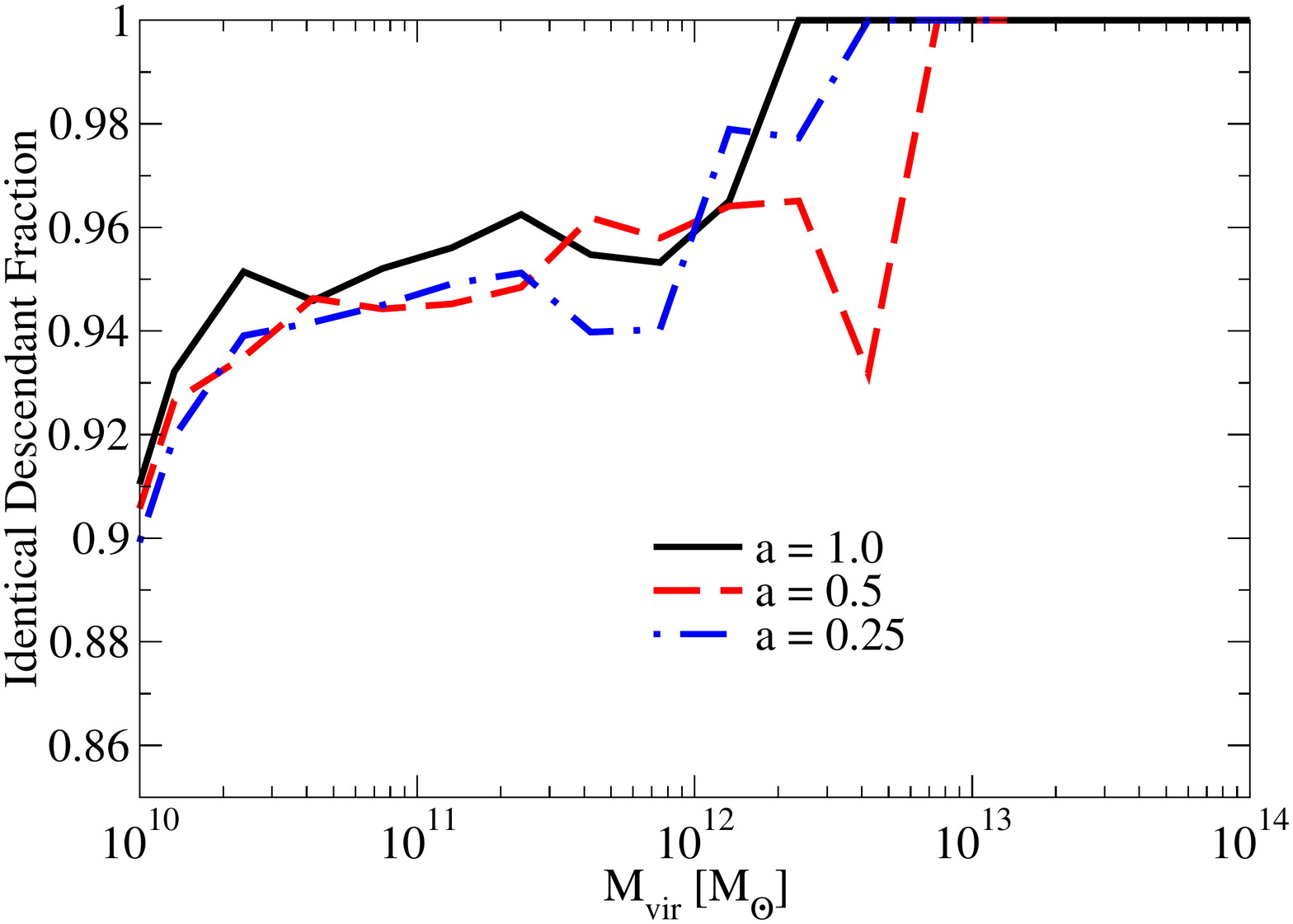}
\caption{Fraction of merging halos for which the merger target calculated via the tidal force method matches the merger target in the particle-based halo merger trees for Bolshoi, using the \textsc{rockstar} halo finder.}
\label{tidal_err}
\end{figure}

\section{Tests of the Method}

\label{s:tests}

\subsection{Tests of Gravitational Evolution}
\label{s:grav_tests}

Despite the complexity of the underlying particle distribution for dark matter halos, the approximation of spherical halos with NFW profiles appears to work remarkably well.  Figure \ref{errors} shows a comparison of the predicted progenitor halo properties with the actual progenitor properties (as determined from particle merger trees).  Figure \ref{errors} demonstrates that the gravitational evolution code can evolve halos from one timestep to another with tightly-controlled positional errors on the order of the force resolution (1 kpc $h^{-1}$) and velocity errors ranging from 2--50 km s$^{-1}$, depending on halo mass, for Bolshoi with the \textsc{rockstar} halo finder.\footnote{See Appendix \ref{a:subfind} for a comparison with the \textsc{subfind} halo finder.}  For \textsc{bdm}, similar results are obtained albeit with higher position and velocity errors by a factor of 1--5, which are detailed in Appendix \ref{s:bdm_r_comp}.  In the Consuelo simulation, we find similar results, with position errors on the order of the force resolution; see Appendix \ref{a:consuelo} for details.

Figure \ref{error_comp} shows the scaling of position and velocity errors with redshift for Bolshoi with the \textsc{rockstar} halo finder.  In Bolshoi, the timestep outputs are not equally spaced; they are between 80--92 Myr before $a = 0.81$ (and 40-46 Myr after that scale \citep{Bolshoi}).  For massive halos, positional errors scale most predominantly as a function of timestep length (suggesting velocity inconsistencies as the most likely reason), whereas for smaller halos, positional errors are fixed closer to the force resolution; these errors result in predicted halo progenitor locations that are well within the virial radius of their actual progenitors across all masses.

Velocity errors for \textsc{rockstar} are somewhat independent of timestep length, and are on the order of 2--30 km s$^{-1}$ at $z=0$---that is, fractional errors on the order of 0.5-3\% in halo peculiar velocities---depending on the mass of the halo.  However, they rise substantially at higher redshifts.  This could be due to a number of effects at high redshift, including higher major merger rates, increasing velocity dispersions, and shorter dynamical times (relative to the timestep length).  For \textsc{bdm}, the velocity errors are elevated (2--100 km s$^{-1}$) independent of redshift, by a factor of 1-5 as compared to \textsc{rockstar} (see Appendix \ref{s:bdm_r_comp}).  These errors suggest either a potential weakness in calculating halo velocities for \textsc{bdm} or a better ability for \textsc{rockstar} to recover halo velocities due to its use of additional phase-space information.

We note some interesting features in the position/velocity errors at low particle counts in Fig.\ \ref{errors}.  For halos with less than 50 particles within the virial radius, it can be very difficult to determine the exact location of the halo density center.  In many cases, \textsc{rockstar} picks the same particles to determine the density center across timesteps.  Due to shot noise in the particle densities, it sometimes picks a different set of particles, leading to a jump in the halo center across timesteps.  This leads to a bifurcation in position errors for low particle counts ($M_h < 10^{10}\Msun$) in Fig.\ \ref{errors}. In all cases, however, the position errors are much smaller than the virial radii of the halos in question.

Another interesting feature is the ``V''-shape in the velocity error distribution for \textsc{rockstar}.  This occurs because \textsc{rockstar} shifts from averaging the mean velocity of particles within $0.1\rvir$ (the ``core velocity'') to averaging all particles within the halo (the ``bulk velocity'') for halos with low particle numbers. As discussed in Appendix \ref{a:subfind} and in \cite{Rockstar}, the core velocity is less self-consistent than the bulk velocity across timesteps; however, using the core velocity gives better consistency with the motion of the halo center across timesteps.  Thus, the velocity errors decrease for moderate (500-particle) to low (100-particle) mass halos, which corresponds to a slight increase in the position errors over that same range.  For very low mass halos ($<$100 particles), the velocity errors increase again due to sampling noise.

\subsection{Tests of Tidal Force Calculations}
\label{s:tidal_tests}
Figure \ref{tidal} shows the maximum tidal fields as calculated for tidally merging halos and non-tidally-merging halos (i.e., most-massive progenitors).  While it is clear that tidal fields above a certain threshold do not necessarily imply a merger, there is a clear threshold in the tidal field below which a merger is very unlikely to happen.  However, this is entirely sufficient for our algorithm to function---as the halos at the next timestep are used to determine the halos at the current timestep which are most-massive progenitors, the remaining halos without descendants at the current timestep must either be tidal mergers or statistical fluctuations, and thus a cut on the magnitude of the tidal field is all that is necessary to distinguish them.

In terms of our ability to predict the target for tidally merging halos (i.e., the halo into which the merging halo dissipates), choosing the halo which exerts the strongest tidal field on the halo in question gives results which are in excellent agreement with particle-based merger trees, as shown in Figure \ref{tidal_err}.  The main disagreements come in cases where a significant fraction of a subhalo's particles are stripped between one timestep and the next, resulting in the subhalo's descendant being assigned to its host halo instead of the remaining subhalo core.  As may be expected, this effect happens more frequently for smaller halos.

\subsection{Tests of Halo Tracking}

\label{s:tracking_tests}

\begin{figure}
\vspace{-5ex}
\plotgrace{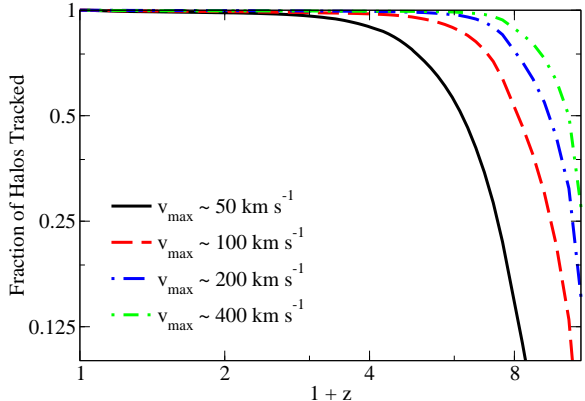}\\[-2ex]
\caption{Fraction of Bolshoi halos at $z=0$ tracked to a given redshift for the \textsc{rockstar} halo finder, as a function of $\vmax$.  Compare to the analogous figure in \cite{Bolshoi}.}
\label{f:anatoly}
\end{figure}

One of the most important tests of any merger tree is the degree to which it correctly follows halo progenitors back in time.  We have imposed stringent cuts on the physicality of halo links, but it is important to show that these cuts do not truncate otherwise correct halo tracks in the merger trees.  We demonstrate our ability to track halos back to early redshifts in Figure \ref{f:anatoly}.  Clearly, halo tracks will end when the most-massive progenitor of a halo falls below the resolution limit of the simulation.  We recover similar tracking statistics at the 50\% threshold as compared to the advanced particle-based trees in \cite{Bolshoi} (e.g., 50\% of $\vmax \sim$ 200 km s$^{-1}$ halos are tracked to $z=7$ in \citealt{Bolshoi}, whereas we track the same fraction halos to $z=8.3$)---these fractions correspond to the expected mass accretion histories of such halos.  However, as compared to \cite{Bolshoi}, we have vastly higher resistance to numerical issues; for example, we track 90\% of $\vmax \sim$ 200 km s$^{-1}$ halos to $z=6.3$, whereas \cite{Bolshoi} tracks the same fraction of $\vmax \sim$ 200 km s$^{-1}$ halos only to $z=1.0$.

\begin{figure}[!h]
\plotgrace{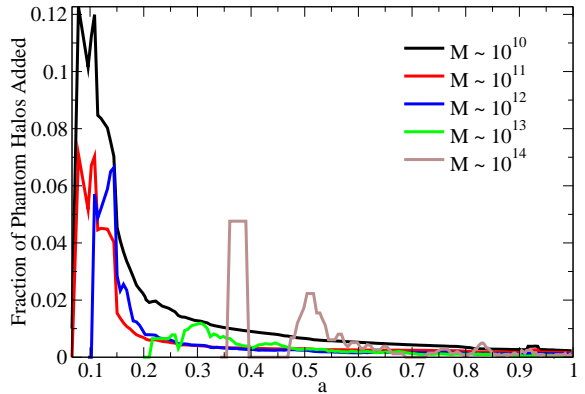}\\[-5ex]
\plotgrace{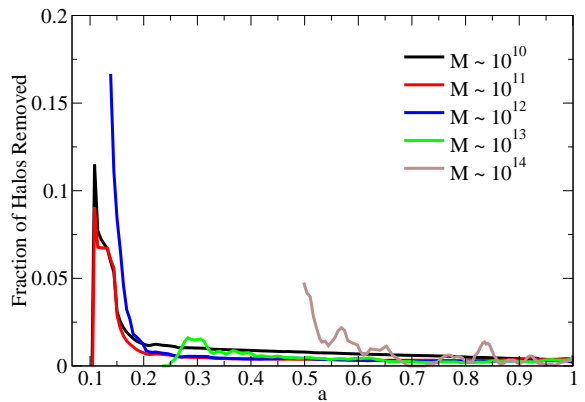}\\[-2ex]
\caption{The fractional changes in the mass function for Bolshoi with the \textsc{rockstar} halo finder for phantom halos added (upper panel) and halos removed (lower panel).  The lower panel excludes phantom halos which were added and later removed; as such, it represents a consistency check for only the halos returned by the halo finder.  The number of halos added by our consistency algorithm is roughly equal to the number of halos deleted, and both are small in comparison to both the total number of halos and the number of subhalos across all masses.  Towards the halo resolution limit, the number of spurious halos and the number of phantom halos necessary to fill in gaps in the merger tracks increases.  All halo masses ($M$) are in units of $\Msun$.}
\label{f:bolshoi_rockstar_consistency}
\end{figure}

\begin{figure}[!h]
\vspace{-5ex}
\plotgrace{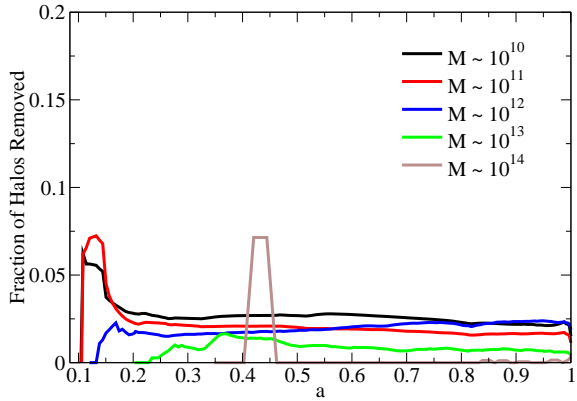}\\[-2ex]
\caption{The fractional changes in the mass function for Bolshoi with the \textsc{bdm} halo finder for halos removed.  As in Fig.\ \ref{f:bolshoi_rockstar_consistency}, this figure excludes phantom halos which were added and later removed; as such, it represents a consistency check for only the halos returned by \textsc{bdm}.  As with \textsc{rockstar}, the number of halos requiring removal is small compared to both the overall mass function and the number of subhalos. All halo masses ($M$) are in units of $\Msun$.}
\label{f:bolshoi_bdm_consistency}
\end{figure}

\subsection{Effects on the Halo Mass Function}

\label{s:mf_effects}

With so many ways to add and delete halos from the merger tree and catalogs, it is important to check the effects on the overall halo mass function.  In the case of \textsc{rockstar} on Bolshoi, the number of added halos (phantoms) and deleted (inconsistent) halos is on the order of 0.5\% for $z<1$ compared to the total mass function; additionally, the number of added and deleted halos are comparable across a wide range of halo masses, as shown in Figure \ref{f:bolshoi_rockstar_consistency}, so that the net effect on the overall mass function is almost negligible.  \textsc{bdm} also performs well; the number of deleted halos is in the 1-2\% range (as shown in Fig.\ \ref{f:bolshoi_bdm_consistency}), with also 1-2\% added phantom halos.

Interpreted in a different way, these numbers imply that the raw halo catalogs from \textsc{rockstar} are internally self-consistent at better than the 1\% level at $z=0$, and nearly as well for \textsc{bdm}.  As discussed in Appendix \ref{a:consuelo}, the results are very similar for the Consuelo simulation, with the \textsc{rockstar} halo finder being self-consistent at the $<0.5$\% level across all halo masses.  While these numbers cannot be directly interpreted as the accuracy of the mass function, they directly represent the \textit{precision} with which each halo finder can recover the mass function.  Nonetheless, the precision does indirectly set a limit for the accuracy of the halo finder at a given timestep; it is worth noting that the precision of both halo finders is significantly better than the current 5\% statistical uncertainties in the halo mass function \citep{tinker-umf}.

We remark that, while the current implementation of our algorithm only allows us to directly test the consistency of positions and velocities, it is in fact possible to use our algorithm to test the accuracy of halo masses and subhalo masses \textit{and correct for systematic bias therein}.  Namely, one can either search directly for systematic alignments of velocity errors (i.e., differences between predicted and actual halo velocities) towards nearby halos, or one can simply adopt a parametrization for the systematic bias in halo masses and search for a fit which minimizes the velocity errors.  We reserve this topic for a future paper, however, as a careful calibration and understanding of the gravitational force formula is necessary to confirm that the acceleration calculations do not suffer from more systematics than the halo mass calculations.

\section{Conclusions and Discussion}
We have developed a new algorithm for creating halo merger trees which explicitly ensures dynamical consistency of halo properties across timesteps.  Our method has several advantages which when combined provide excellent robustness and tracking of both halos and subhalos:
\begin{enumerate}
\item The ability to more accurately track halos than particle-based merger trees.
\item The ability to explicitly evaluate the precision with which the halo finder can recover halo positions and velocities.
\item The ability to correct for halo finder incompleteness by adding halos with gravitationally consistent properties to the halo catalogs.
\item The ability to correct for incorrect tidal mergers in particle-based trees.
\item The ability to construct merger trees even in the absence of full particle tracking; in addition, the ability to construct merger trees even when particle-based merger trees are available only for a small region of the simulation.
\item The ability to remove halos which fail any of a large number of sanity tests (gravitational inconsistency, tidal inconsistency, tracking inconsistencies) to increase the purity of the resulting merger trees.  In addition, the ability to uncover previously unknown problems in halo finders.
\item The ability to explicitly evaluate the self-consistency of the mass function returned by the halo finder.  In the future, the ability to explicitly evaluate the accuracy of the mass function returned by the halo finder.
\end{enumerate}

The code used is publicly available at {\tt http://code.google.com/p/consistent-trees}.  This algorithm has been used to create merger trees for two simulations, the Bolshoi simulation ($2048^3$ particles in a 250 \hmpc{} box) and the Consuelo simulation ($1400^3$ particles in a 420 \hmpc{} box), and have additionally compared two halo finders (\textsc{bdm} and \textsc{rockstar}) on the Bolshoi simulation.  The halo finders perform similarly for many instances, and self-consistency is problematic only at the 1-2\% level for both.

The defining feature of our algorithm is the ability to predict the evolution of halo locations, velocities, and properties.  This ability applies \textit{within} timesteps as well, which has relevance for semi-analytical models (of, e.g., reionization) which may require information about halo properties at many more timesteps than are storable from the simulation.  Our approach allows for effectively infinite timestep resolution in terms of individual halo properties; in addition, because tidal forces are calculated in each step, it becomes possible to estimate the timing of halo mergers and thereby recover all the information which is lost by saving fewer snapshots from a simulation.

The merger trees and halo catalogs thus generated are useful not only for improved cosmological predictions from simulations, but because of the improved tracking performance, they are also useful for precision predictions of a wide range of observables related to merger trees: the galaxy-galaxy merger rate, the dynamical friction timescale of subhalos, halo mass accretion histories (especially as functions of environment and assembly time), and semi-analytical / semi-empirical models of galaxy formation; several studies exploring these predictions are already in progress.  

\label{s:conclusions}

\acknowledgements 
This research was supported by a NASA HST Theory grant HST-AR-12159.01-A, by the National Science Foundation under grant NSF AST-0908883, and by the U.S. Department of Energy under contract number DE-AC02-76SF00515.  We would like to thank Darren Croton, Rachel Somerville, Yu Lu, Oliver Hahn, Tom Abel, Markus Haider, and Joanne Cohn for useful discussions.  We also thank the LasDamas Collaboration for input on the Conseulo simulation, which was run on the Orange cluster at SLAC.  The Bolshoi simulation was run at the NASA Ames Research Center.  We gratefully acknowledge the support of Stuart Marshall and the SLAC computational team, as well as the computational resources at SLAC.

\newpage 

\bibliography{tree}

\appendix

\section{\textsc{bdm} position / velocity tracking as compared to \textsc{rockstar}}

\label{s:bdm_r_comp}

\begin{figure}[!h]
\plotminigrace{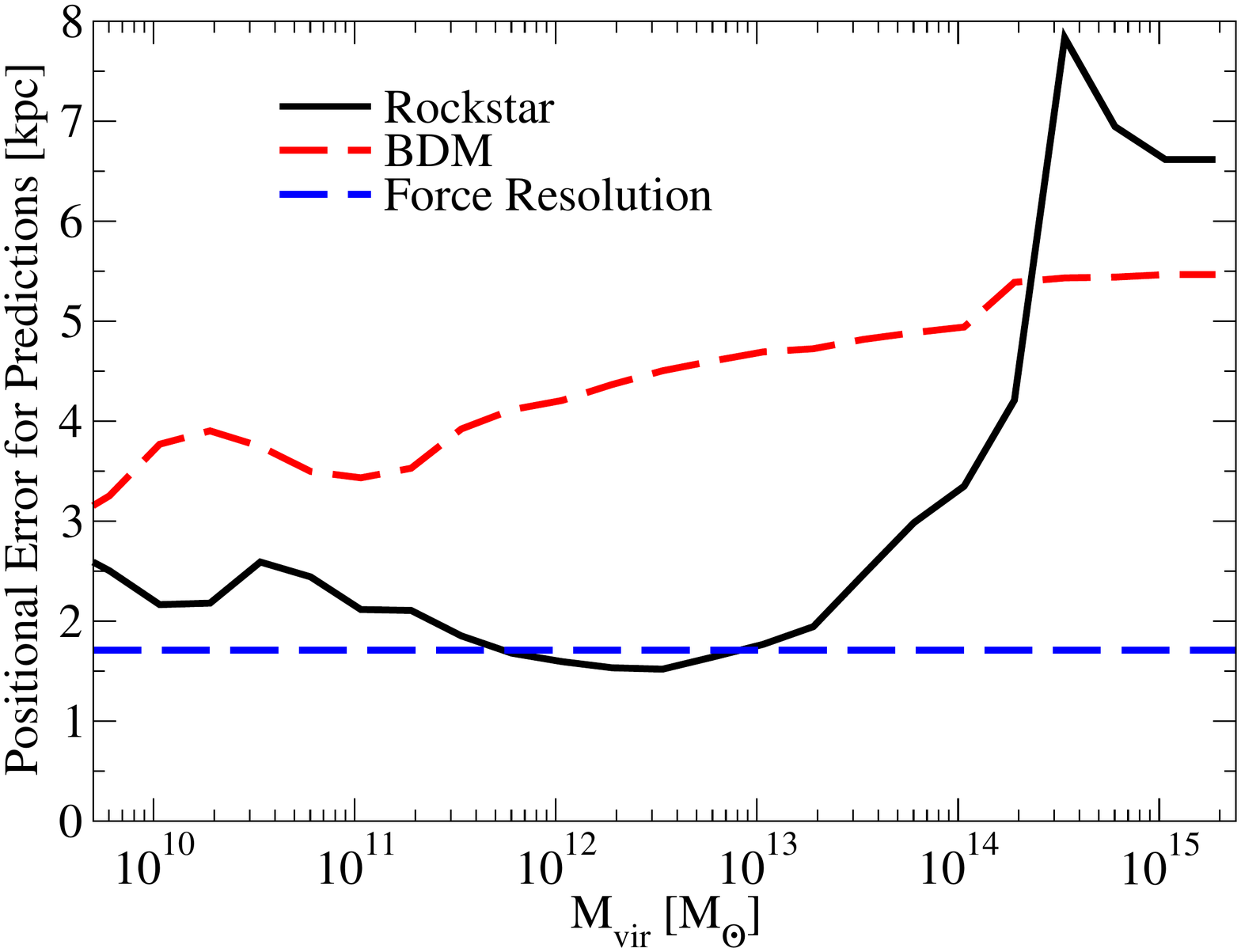}
\plotminigrace{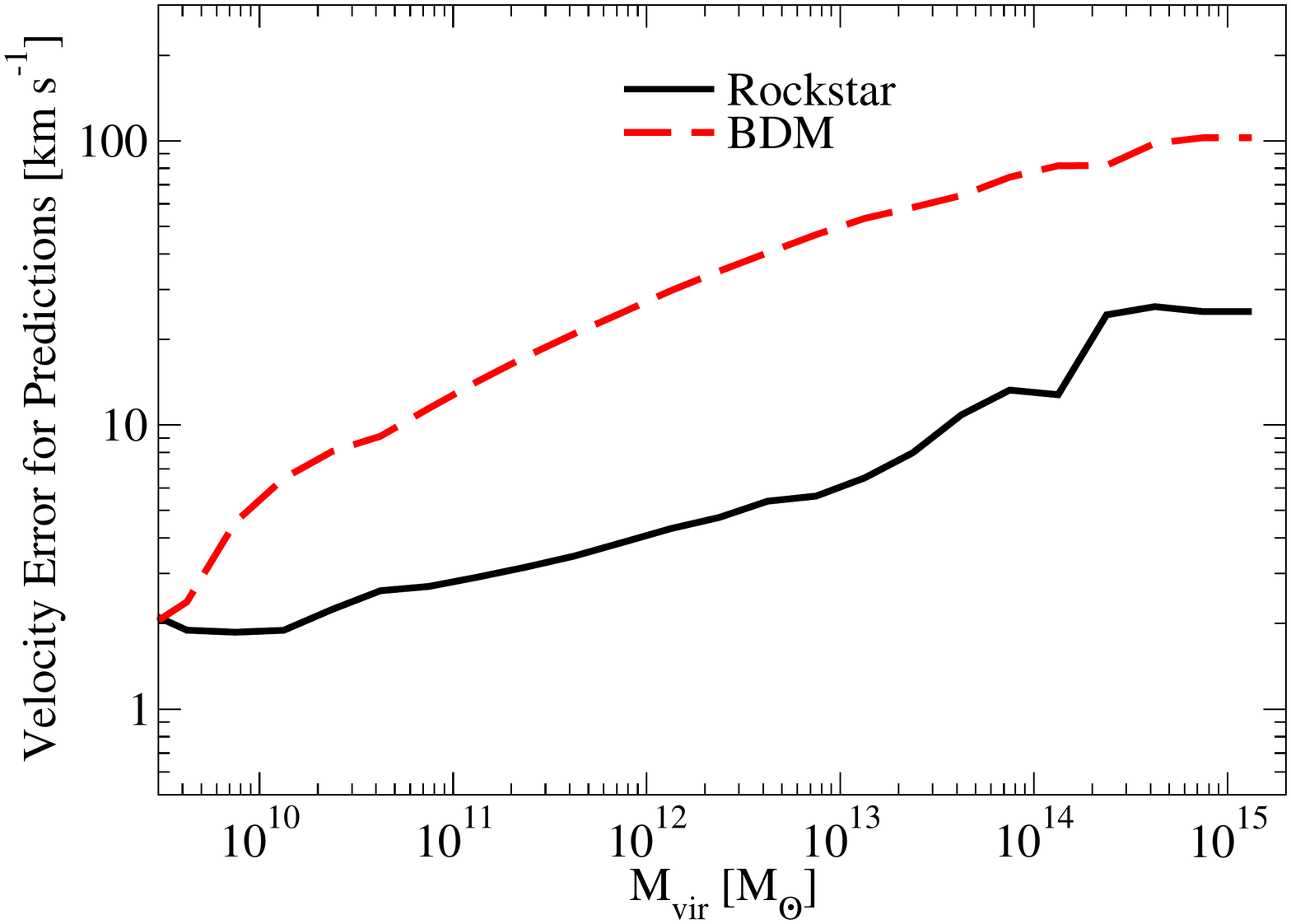}
\caption{Comparison of position and velocity errors at $z=0$ for the \textsc{rockstar} and \textsc{bdm} halo finders on Bolshoi, similar to Fig.\ \ref{errors}.}
\label{f:bdm_rockstar_comp}
\end{figure}

Figure \ref{f:bdm_rockstar_comp} shows position and velocity tracking errors for \textsc{bdm} as compared to \textsc{rockstar}.   For low-mass halos ($M < 10^{13} \Msun$), \textsc{rockstar} performs ideally, almost at the force resolution of the simulation.  In comparison, \textsc{bdm} gives positions accurate to within a factor of a few of the force resolution (2-3).  For high-mass halos ($M > 10^{13} \Msun$), both halo finders perform similarly.  Because of its ability to find halos in phase space, \textsc{rockstar} is always better able to recover halo velocities than \textsc{bdm}; the resulting velocity errors are smaller by a factor of 1-5.

\section{Consuelo}
\label{a:consuelo}

\begin{figure}[!ht]
\vspace{-5ex}
\plotminigrace{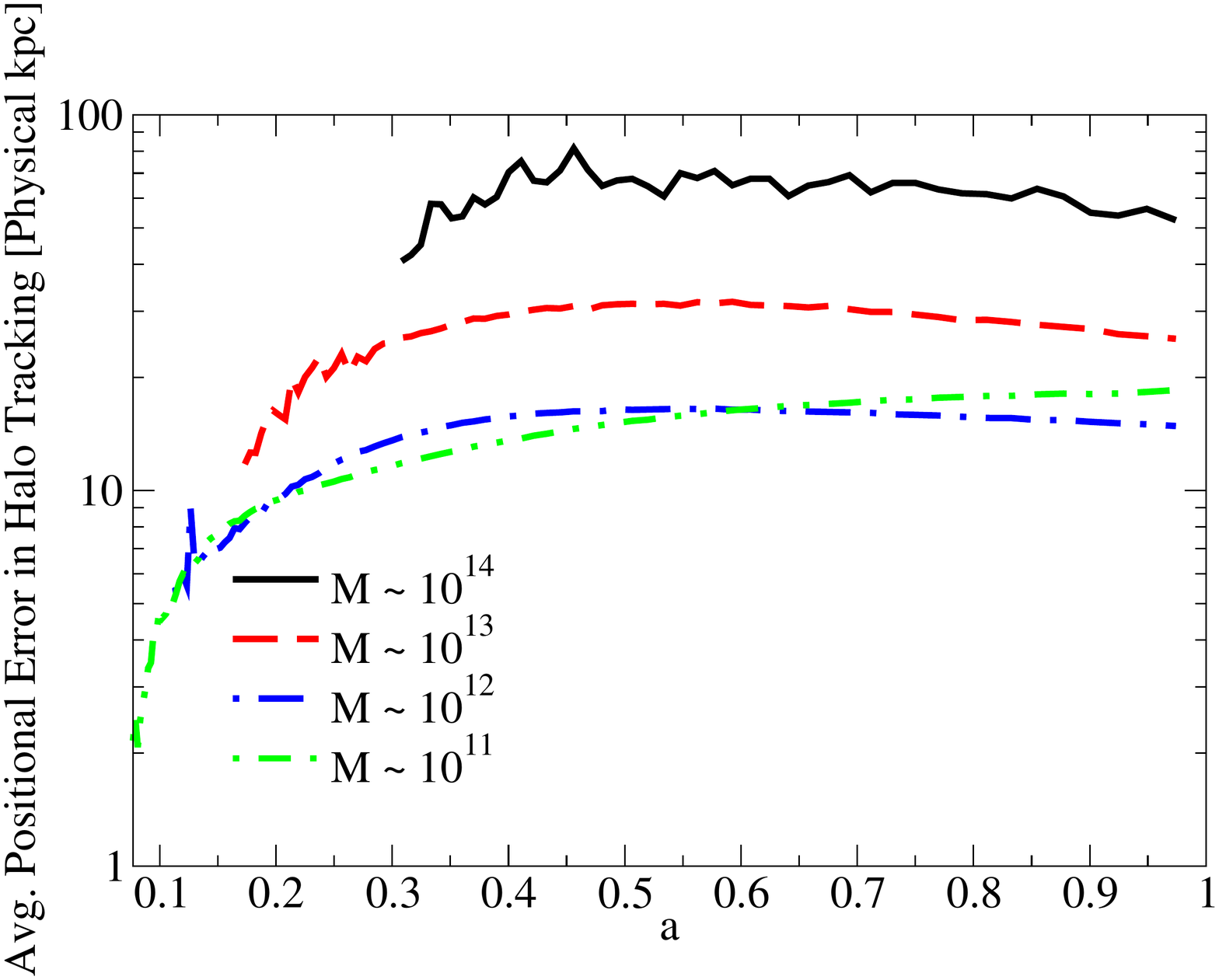}
\plotminigrace{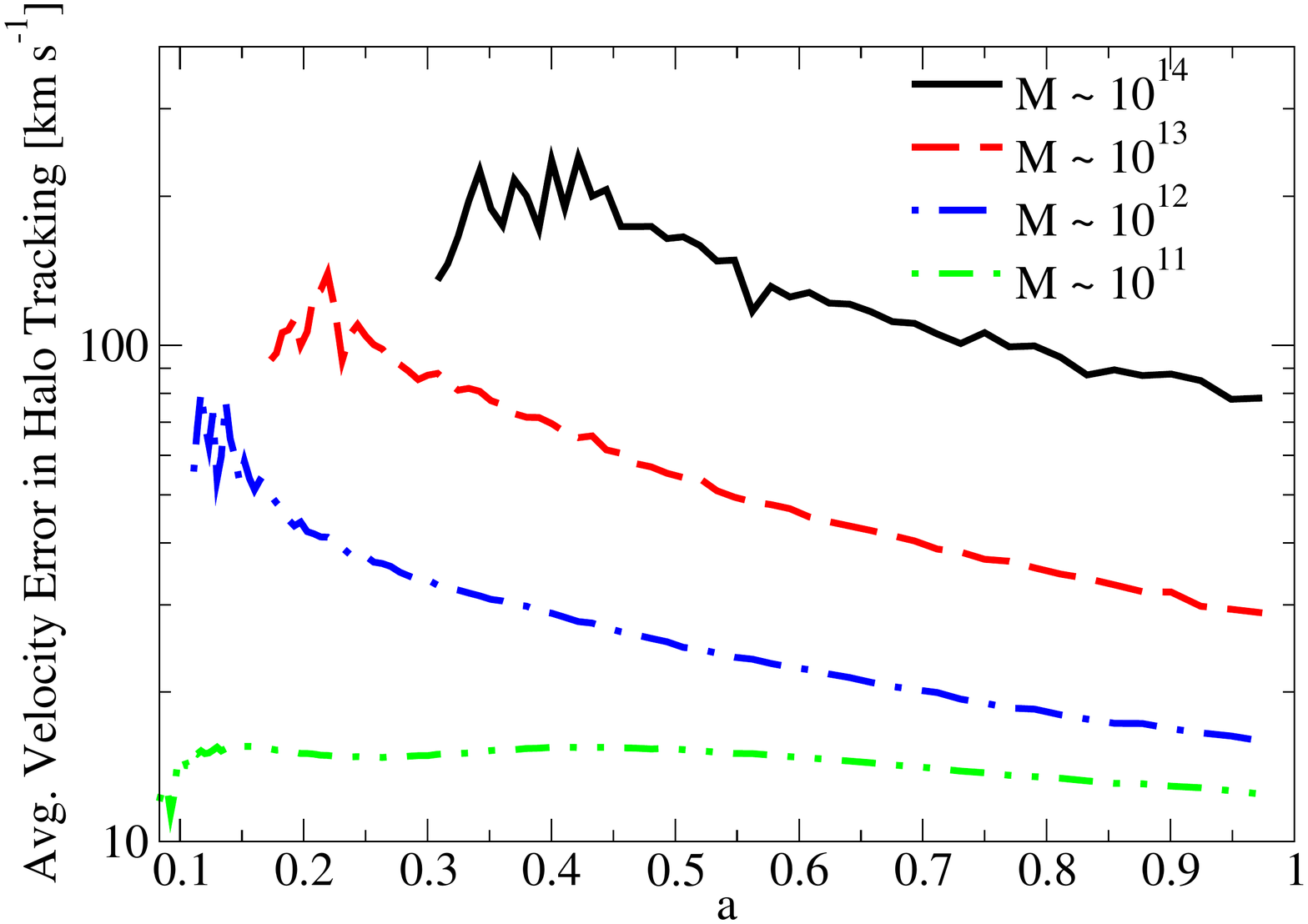}
\caption{Comparison of errors at different timesteps for the \textsc{rockstar} halo finder on Consuelo, analogous to Fig.\ \ref{error_comp}.}
\label{error_comp_consuelo}
\end{figure}

\begin{figure}[!ht]
\vspace{-5ex}
\plotminigrace{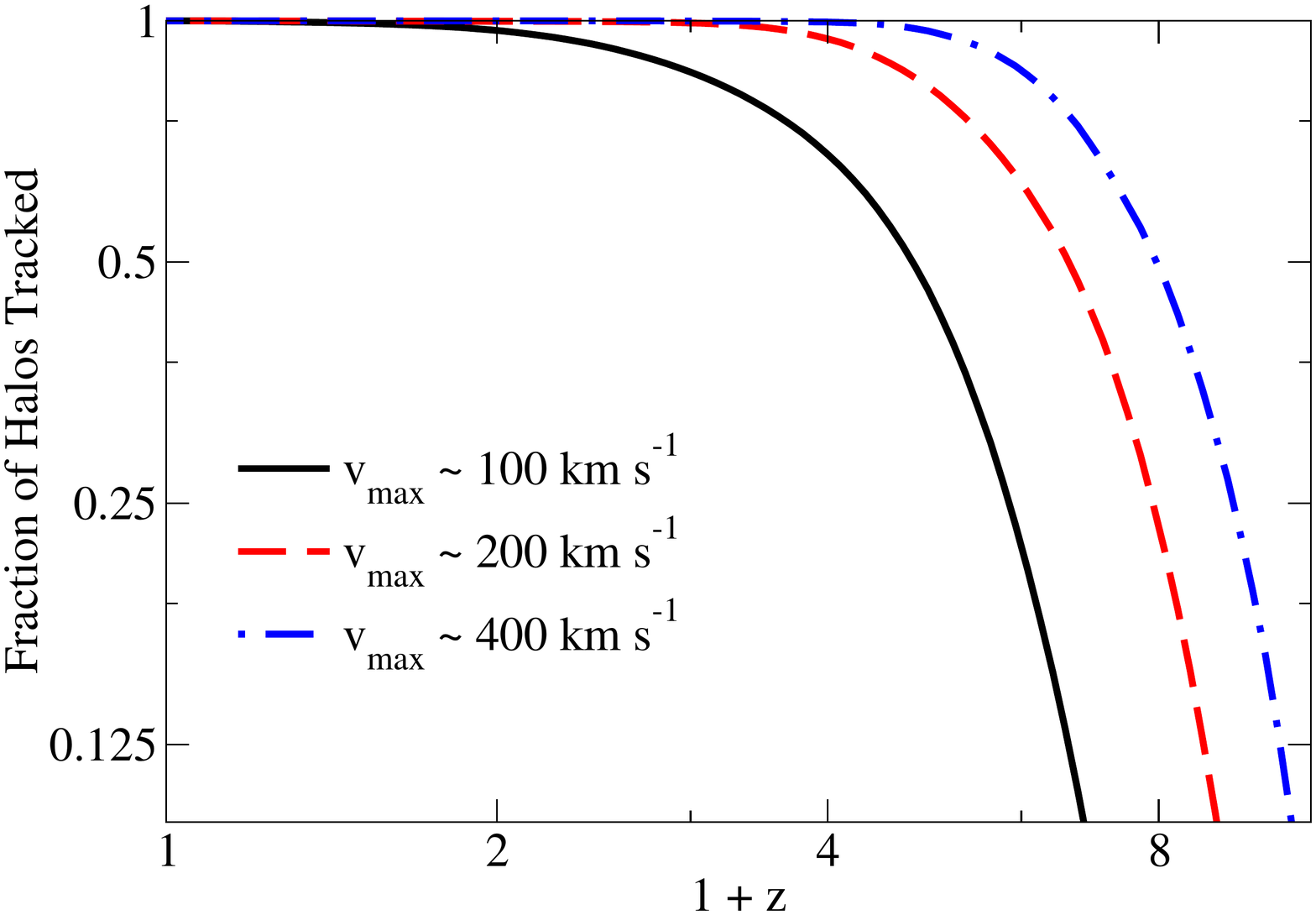}\plotminigrace{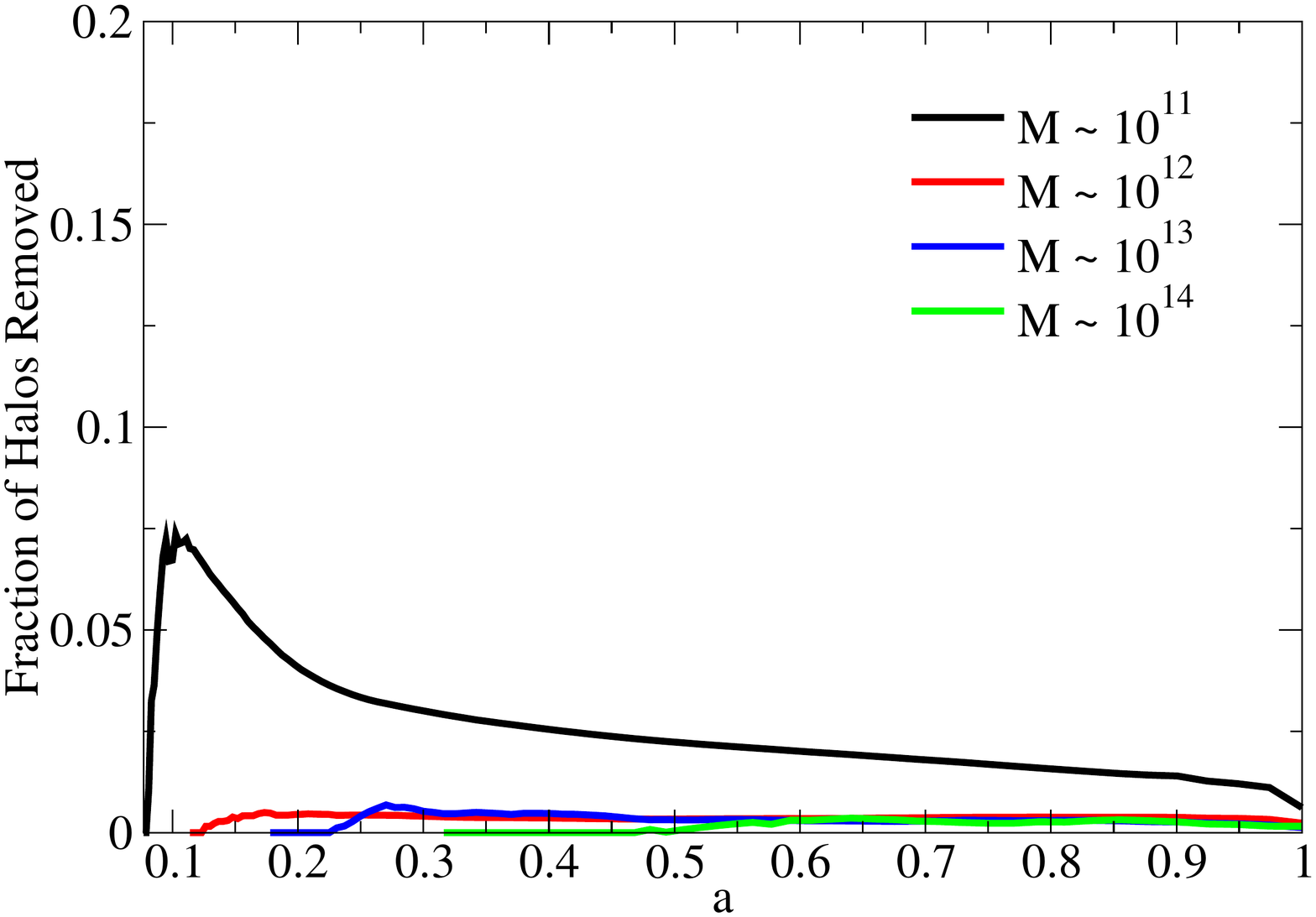}
\caption{\textbf{Left panel:} Fraction of Consuelo halos at $z=0$ tracked to a given redshift for the \textsc{rockstar} halo finder, as a function of $\vmax$, analogous to Fig.\ \ref{f:anatoly}.  \textbf{Right panel:} Fraction of halos found by the \textsc{rockstar} halo finder which were removed in the process of physical consistency checking as a function of mass and redshift, analogous to Figs.\ \ref{f:bolshoi_rockstar_consistency} and \ref{f:bolshoi_bdm_consistency}.}
\label{f:anatoly_consuelo}
\end{figure}

For comparison with Bolshoi, we include results for the Consuelo simulation analogous to Fig.\ \ref{error_comp} in Fig.\ \ref{error_comp_consuelo} and figures analogous to Figs.\ \ref{f:anatoly}, \ref{f:bolshoi_rockstar_consistency}, and \ref{f:bolshoi_bdm_consistency} in Fig.\ \ref{f:anatoly_consuelo}.  We find identical results for the Consuelo simulation as compared to the Bolshoi simulation, with only a few exceptions.  The main difference in terms of the more limited mass and force resolution of Consuelo means that the positional errors for recovered halos are higher (Fig.\ \ref{error_comp_consuelo}) and that halos with maximum circular velocities of 50-100 km s$^{-1}$ are no longer above the resolution limit (and hence are excluded from Fig.\ \ref{f:anatoly_consuelo}).  In addition, the timesteps are logarithmically spaced in scale factor, meaning that at earlier times, the timesteps are more closely spaced.  This contributes to the positional errors becoming smaller with increasing redshift for Consuelo (as contrasted with them becoming larger with increasing redshift in Bolshoi).

\section{\textsc{subfind}}
\label{a:subfind}

\begin{figure}[!h]
\plotminigrace{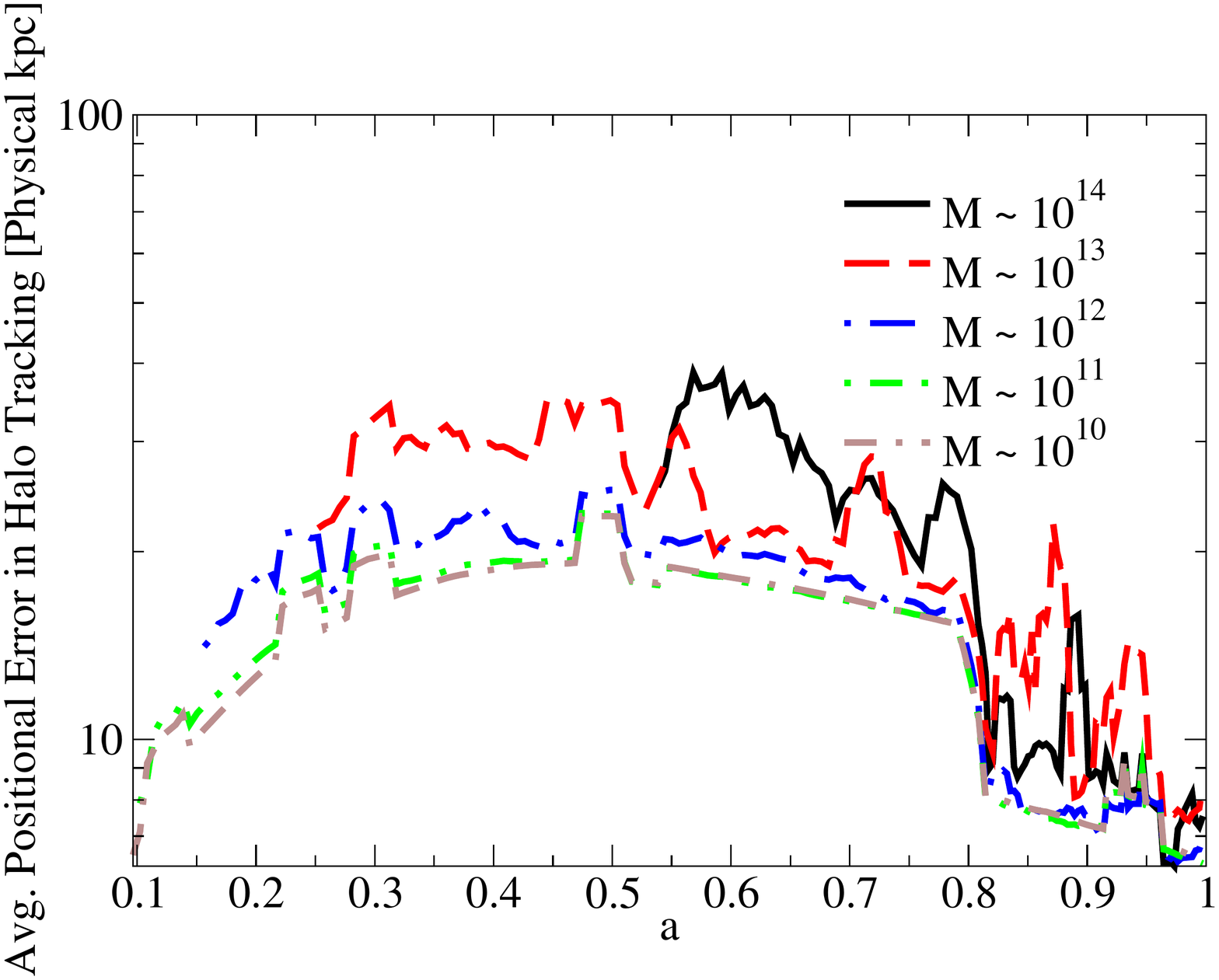}
\plotminigrace{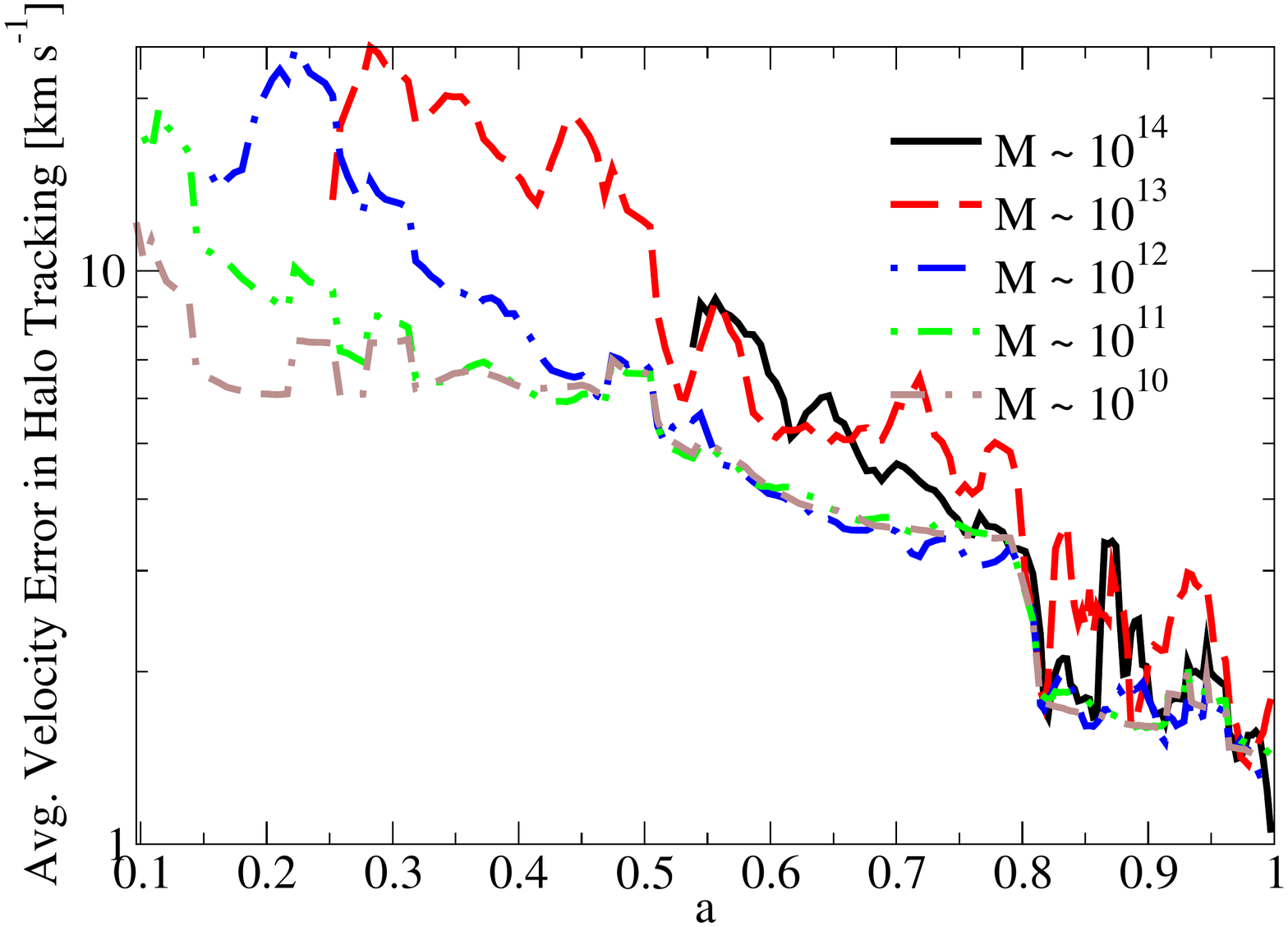}
\caption{Comparison of errors at different timesteps for the \textsc{subfind} halo finder on a subregion of Bolshoi, analogous to Fig.\ \ref{error_comp}.}
\label{error_comp_subfind}
\end{figure}

We have additionally used our algorithm on the \textsc{subfind} halo finder \citep{Springel01} on a small subvolume (50 Mpc $h^{-1}$ on a side) of Bolshoi.  Because the algorithm described in this paper requires halo masses even for subhalos, and because \textsc{subfind} by default only returns particle membership for subhalos, a spherical overdensity mass calculator was used on the halo centers returned by \textsc{subfind} to calculate the unbound mass within $R_\mathrm{vir}$ for host halos as well as $\vmax$ and $R_{\vmax}$ for all halos; these latter two properties were used to estimate subhalo mass.  Because the mass derived this way is inconsistent with the mass for host halos, we do not show results for halo tracking or self-consistency, which would unfairly penalize \textsc{subfind}.

Nonetheless, it is possible to calculate the position / velocity precision for halos returned by \textsc{subfind}, as the number of halos affected by inconsistent subhalo masses across a single timestep is small.  We show results analogous to Fig.\ \ref{error_comp} in Fig.\ \ref{error_comp_subfind}.  We find that \textsc{subfind} appears to have exquisite velocity precision at the expense of some position precision (2-5 times worse than \textsc{rockstar}; see Fig.\ \ref{error_comp_subfind}).  \textsc{subfind}'s velocity precision does not necessarily translate into velocity \textit{accuracy}, however.  Indeed, \textsc{subfind} averages particle positions to yield a bulk halo velocity, but as demonstrated in \cite{Rockstar}, the difference between the halo core velocity (which would correspond more closely to the central galaxy velocity) and the halo bulk velocity can be on the order of 20\% of the halo velocity dispersion, up to 400 km s$^{-1}$ for the largest clusters at $z=0$.  Hence, while \textsc{subfind}'s velocities are remarkably consistent, they do not necessarily correspond to observable properties.

\end{document}